\documentclass[12pt,letterpaper]{article}
\usepackage[a4paper, total={7in, 10in}]{geometry}

\usepackage{graphicx}
\usepackage{helvet}
\usepackage{authblk}
\usepackage{hyperref}
\usepackage{amsmath} 
\usepackage{amssymb} 
\usepackage{orcidlink}
\usepackage{amsthm, amssymb, amsfonts}
\usepackage[super,comma,sort&compress]  
   {natbib}
\usepackage{booktabs}
\usepackage{nameref}
\usepackage{amsthm}
\usepackage{float}

\makeatletter
\renewcommand{\maketitle}{\bgroup\setlength{\parindent}{0pt}
\begin{flushleft}
  \textbf{\@title}
  
  \@author
\end{flushleft}\egroup}
\makeatother

\title{Grid-Supporting Equipment Supply Chains Constrain the Feasible Pace of Power System Expansion}

\date{}

\author[1,3,**]{Boyu Yao}
\author[1,2,3,*]{Yury Dvorkin}

\affil[1]{Department of Civil and Systems Engineering, Johns Hopkins University, Baltimore, MD, USA}
\affil[2]{Department of Electrical and Computer Engineering, Johns Hopkins University, Baltimore, MD, USA}
\affil[3]{Ralph O’Connor Sustainable Energy Institute, Johns Hopkins University, Baltimore, MD, USA}

\affil[*]{Correspondence: ydvorki1@jhu.edu}
\affil[**]{Correspondence: byao3@jhu.edu}

\begin{document}

\maketitle

\section*{CONTEXT \& SCALE}

Power system expansion explicitly considers generation, storage, transmission, and demand, while the equipment required to connect, convert, regulate, and condition electricity is often assumed to be readily available. This class of grid-supporting equipment (GSE), including transformers, converters, inverters, power conversion systems, and uninterruptible power supplies, forms the physical interface linking supply- and demand-side electrification. Rapid growth in data centers, electrified industry, electric vehicles, and inverter-based generation is increasing demand for GSE. Because these assets rely on specialized manufacturing processes and critical material inputs, their availability may constrain the feasible pace of grid expansion.

This study integrates engineering-based material accounting and power system planning to examine GSE as a deployment-constrained component of power system expansion. First, we develop a dynamic stock-and-flow model that links generation and load growth projections to GSE requirements and accounts for aging and replacement of existing equipment. Second, we compile equipment-level bills of materials (BOMs) and normalize material quantities by apparent-power capacity to derive material-intensity coefficients in ton/GVA, which convert GSE capacity requirements into physical material demand. Third, we use global multi-regional supply-use tables (MRSUTs) to characterize the upstream production and regional sourcing structure associated with GSE-related material supply. This enables identifying cross-sectoral dependencies, geographic concentration, and supply bottlenecks for each material. Finally, we apply these material constraints in a GSE deployment model to quantify unmet GSE demand across growth, trade disruption, and mitigation scenarios.

\section*{HIGHLIGHTS}

\begin{itemize}
\item GSE links generation, transmission, storage, and demand from data centers and EVs.
\item Critical materials provide a physically grounded proxy for GSE deployment constraints.
\item High load growth yields 269.6--274.1 GVA (28.5\%--28.6\%) U.S. GSE shortages by 2030.
\item Copper binds first; steel and nickel bottlenecks worsen under trade disruption.
\item Grid-enhancing technologies provide only partial relief from GSE shortages.
\end{itemize}

\section*{SUMMARY}

Power system expansion depends on the equipment required to connect, convert, regulate, and condition electricity, yet GSE is rarely modeled as an explicit constraint. We develop a framework integrating dynamic stock-flow modeling, BOM compilation and material intensity derivation, multi-regional supply-use analysis, and GSE deployment modeling to quantify GSE deployment requirements and upstream material dependence. Because manufacturing information is often fragmented or proprietary, we use critical material availability as physical proxy of GSE supply constraints. In a U.S. case study, GSE shortages reach 269.6--274.1 GVA (28.5\%--28.6\%) by 2030 under high-growth conditions. Copper is the first material to limit GSE availability from 2027 onward, while steel and nickel become additional constraints later. Trade disruption intensifies shortages, while dynamic transformer rating provides targeted but partial relief. These results show that grid expansion depends on the timely manufacturability, replacement, and material support of GSE, motivating planning frameworks that explicitly incorporate deliverability, supply chain exposure, and resilience strategies.

\section*{GRAPHICAL ABSTRACT}
\begin{figure}[H]
    \centering
    \includegraphics[width=\linewidth]{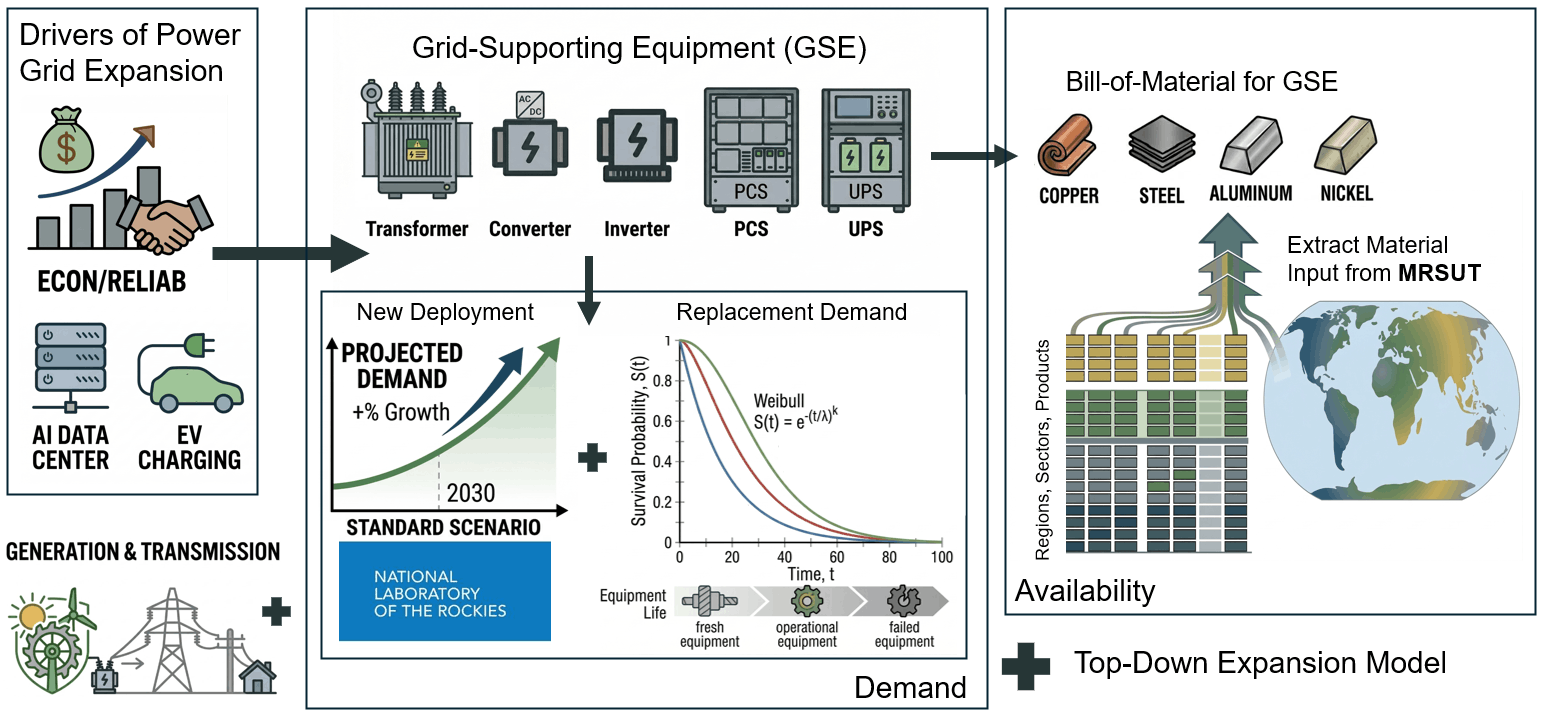}
\end{figure}

\section*{KEYWORDS}

Grid-supporting equipment, Critical material, Supply chain, Power system resilience, Infrastructure bottlenecks

\newpage

\section*{INTRODUCTION}

\begin{figure}[b!]
    \centering
    \includegraphics[width=\linewidth]{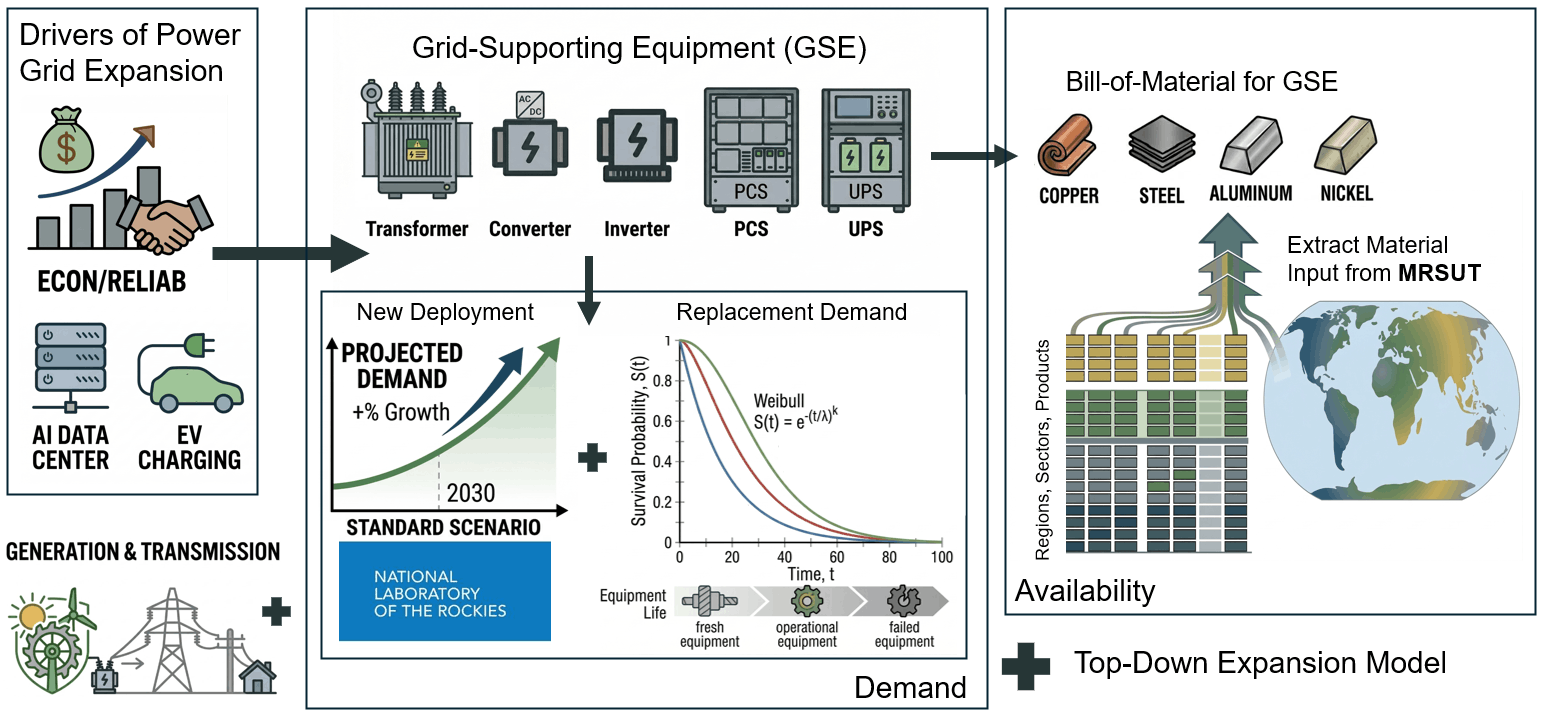}
    \caption{\textbf{Conceptual framework of the study.} Grid-expansion drivers determine demand for GSE, whose new deployment and replacement needs are linked to BOM-derived material intensities and upstream material availability through MRSUT-based supply chain tracing. These components jointly inform the top-down GSE deployment model.}
    \label{fig:1}
\end{figure}

Power system expansion is driven by economic activity and the need to maintain system reliability \cite{doe2025, burke2018impact}. Presently, electricity-intensive industries such as artificial intelligence (AI)-oriented data centers, automated manufacturing, and other electrified end uses are growing faster than the average economic growth that power system planners anticipated; for example, data centers alone could consume 9\%--17\% of U.S. electricity generation by 2030 \cite{epri_powering_intelligence_2026}, thereby increasing demand for the equipment required to connect, convert, and condition electricity across the grid \cite{lin2024exploding, shehabi20242024}. We refer to this class of assets as grid-supporting equipment (GSE), including bulk power transformers, solar photovoltaic (PV) inverters, wind turbine converters, battery storage power conversion systems (PCS), data center transformers and data center uninterruptible power supplies (UPS), and EV charger PCS. The GSE availability, in turn, depends on manufacturing systems and material supply chains that are already under strain, with reported shortages, long lead times, and increasing exposure to constrained critical inputs \cite{nguyen2022electric, mckenna2024major}. Thus, power system expansion may not only be constrained by generation and transmission assets, but also by the GSE availability. If GSE supply lags, expansion pathways such as AI infrastructure buildout may not be feasible in practice and cause deployment delays.

Understanding these bottlenecks requires recognizing GSE's functional role to \textit{transform} power (change voltage levels for transmission and end use), \textit{regulate} power (control power flow, voltage, and system stability), and \textit{condition} power (maintain waveform quality and compatibility for reliable operation) \cite{blaabjerg2004power}. Supply-side GSE, including generator step-up (GSU), transmission, and distribution transformers, serves as the essential electromagnetic interface for primary power exchange, bulk regional transfer, and localized voltage regulation \cite{nguyen2022electric}. At the same time, the integration of inverter-based resources (IBRs), such as solar, wind, and battery storage, depends on supply-side GSE, including inverters, converters, and PCS, to deliver grid-compatible alternating current (AC) and support controlled operation \cite{eltawil2010grid, blaabjerg2011power, wang2016review}. On the demand side, load-side GSE must meet the distinct power conditioning requirements of emerging high-density consumers. Data centers rely on dedicated transformers and UPS to ensure highly reliable power quality for digital workloads, while EV infrastructure requires advanced PCS to manage high-capacity charging demand \cite{karpati2012uninterruptible, safayatullah2022comprehensive}. Because GSE forms the physical interface between electricity supply, transmission and distribution, and end use, its deployment is required across both supply-side expansion and emerging electricity demand. Additionally, annual GSE requirements are shaped by the need to replace existing legacy equipment. The amount of equipment that must be manufactured can therefore grow differently from the net capacity additions represented in power system expansion scenarios.

GSE supply is still commonly treated implicitly in power system planning. Several related research streams have examined the material, supply chain, and infrastructure constraints associated with power system expansion. Studies of critical material requirements have quantified how the expansion of wind, solar, storage, and other energy technologies increases demand for metals and minerals \cite{wang2023future, dong2025sub, yao2025understanding, hou2026mineral}. Building on this work, material constraints have increasingly been incorporated into system-level models, although through different approaches. Integrated Assessment Models (IAMs) examine how material availability may reshape long-term energy transition pathways \cite{qiu2024impacts}, while Capacity Expansion Models (CEMs) can impose material availability, costs, manufacturing, or deployment constraints directly on power-sector investment decisions \cite{brown2025no, yao2026generation}. Computable General Equilibrium (CGE) frameworks provide a complementary economy-wide perspective by factoring in the effects of cross-sectoral competition, trade, and price responses as energy and material demands change \cite{hotchkiss2024comparing}. In these modeling approaches, it is common for material requirements and supply constraints to be linked to primary generation or storage technologies. However, the transformers and power electronic equipment required to connect and operate this capacity remain rarely accounted for, including the replacement needs for legacy equipment. Consequently, these approaches can reveal how material scarcity changes transition pathways or technology deployment, but not whether shortages arise in the interfacing equipment itself.

On the other hand, a related body of supply chain resilience literature has documented manufacturing concentration, material and component dependencies, import exposure, and other supply vulnerabilities in transformer, solar inverter, and wind equipment supply chains \cite{nguyen2022electric, barlow2024analysis, baranowski2022wind}. These studies provide detailed evidence on where equipment-specific supply risks arise, but generally do not translate those vulnerabilities into the GSE capacity that can be deployed. In parallel, power system planning studies have emphasized that infrastructure availability, including transmission capacity and interconnection, can shape feasible expansion pathways \cite{gorman2025grid, zhu2026integrated}. This literature establishes that planned generation capacity alone does not determine whether system growth can be realized, but largely focuses on transmission capacity, network expansion, and interconnection constraints rather than on the availability of the transformers and power electronic equipment needed to support that growth. Our study connects energy system  and material-constrained supply chain modeling by explicitly representing GSE between projected generation and load growth and upstream material supply. We relate the new deployment and replacement requirements to equipment-specific material demand and regional sourcing, and translate upstream material constraints into GSE deployment and shortage outcomes. This allows us to assess whether the supporting equipment implied by a projected expansion pathway can be supplied at the required pace.

The practical relevance of this gap is already visible in the electric power sector, where transformer lead times have increased up to 2 years (a fourfold increase on pre-2022 lead times) \cite{mckenna2024major}. Rapid power system growth is also placing additional pressure on the materials and manufacturing systems that support GSE production. In Europe, planned deployment of 429 transformers by 2030 is expected to substantially increase demand for key inputs, with copper emerging as an important constraint; in 2030 alone, grid expansion could require 324~kt of copper \cite{JRC143190}. Comparable concerns are arising in the United States, where industry assessments indicate that existing vulnerabilities in transformer manufacturing, critical material inputs, and specialized labor may be further worsened by trade restrictions on precision-engineered components, increasing costs and potentially slowing grid modernization \cite{NEMA2025Section232}. Together, these developments indicate that GSE shortages are not only a prospective modeling concern, but are already emerging through longer procurement times and growing pressure on material and manufacturing supply chains.

Modeling GSE constraints is challenging for several reasons. First, GSE is highly heterogeneous. Unlike more modular clean energy technologies, such as solar PV panels and lithium-ion batteries, transformers and power electronic systems vary substantially across voltage classes, cooling configurations, control architectures, and jurisdiction-specific engineering requirements \cite{nguyen2022electric}. Although efforts are underway to standardize transformer components through modular design \cite{doe2024lptresilience, 10975224}, much of the current manufacturing landscape remains only weakly standardized. Second, future GSE demand cannot be inferred from new deployment alone, because the installed fleet is aging and will require substantial replacement over time. Third, detailed information on manufacturing throughput, subcomponent sourcing, and assembly capacity is often proprietary or unavailable in public datasets \cite{yang2016building}. These features make direct representation of GSE manufacturing constraints difficult at the system level.

We therefore use the critical materials embodied in GSE as a tractable and physically grounded representation of one important dimension of GSE supply constraints. In the absence of consistent equipment-level manufacturing data, we estimate GSE material availability by combining specific equipment material requirements with economy-wide data on how those materials are used in manufacturing and sourced across regions \cite{olivetti2017lithium, hache2019critical, stadler_2025_15689391}. This representation accounts for the material flows supporting GSE-related production and their regional sourcing patterns, which together identify the upstream material and sourcing constraints as a factor for GSE deployment.

This study integrates engineering-based material accounting and power system planning to examine GSE as a deployment-constrained component of power system expansion. We first connect generation and load growth projections to the GSE needed to support that growth, while accounting for both new deployment and replacement of existing equipment. We develop a dynamic stock-and-flow model to represent the aging of both supply-side and load-side GSE and to project associated replacement needs. Second, we compile equipment-level bills of materials (BOMs), which characterize the quantities and composition of constituent materials in each GSE class, and normalize these data by equipment apparent-power capacity, expressed in gigavolt-amperes (GVA), to derive material-intensity coefficients in ton/GVA. These coefficients convert projected GSE capacity requirements into physical material demand. Third, we combine these engineering estimates with global multi-regional supply-use tables (MRSUTs) to trace the upstream supply chains associated with the materials required for U.S. GSE and identify cross-sectoral dependencies, geographic concentration, and material bottlenecks. Finally, we use these material constraints in a GSE deployment model to quantify how much required equipment can be supplied and where unmet GSE demand emerges under different load growth, electrification, trade disruption, and mitigation scenarios.

\section*{RESULTS}
\label{sec:result}

\begin{figure}[b!]
    \centering
    \includegraphics[width=\linewidth]{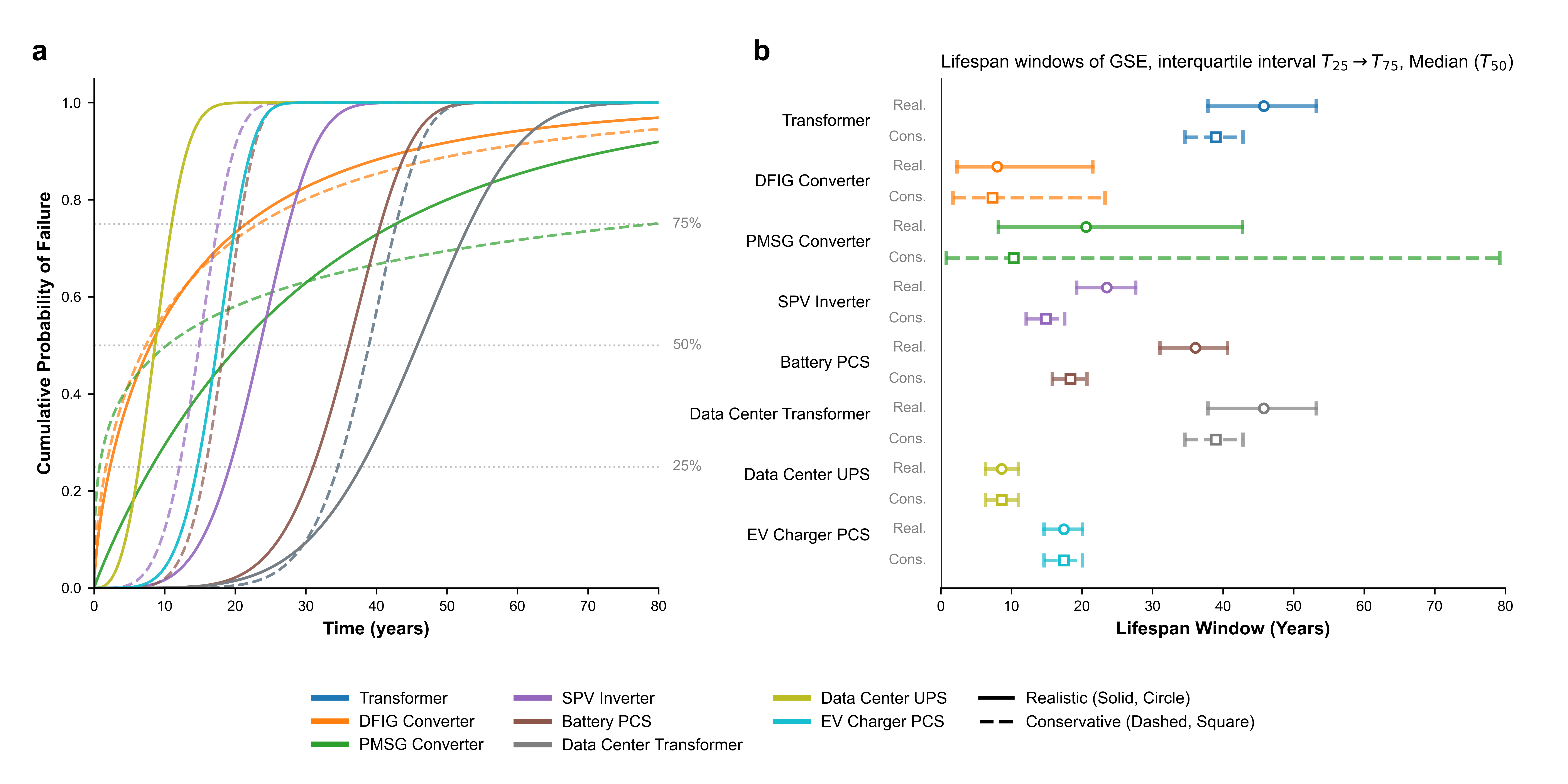}
    \caption{\textbf{Weibull-distributed lifetime characterization of GSE with optimistic and pessimistic lifetimes.}
(a) Cumulative failure probability distributions for GSE with optimistic and pessimistic operation. (b) Interquartile lifespan windows ($T_{25}\rightarrow T_{75}$) and median lifetimes ($T_{50}$) for each equipment type.}

    \label{fig:2_1}
\end{figure}

\begin{figure}[b!]
    \centering
    \includegraphics[width=\linewidth]{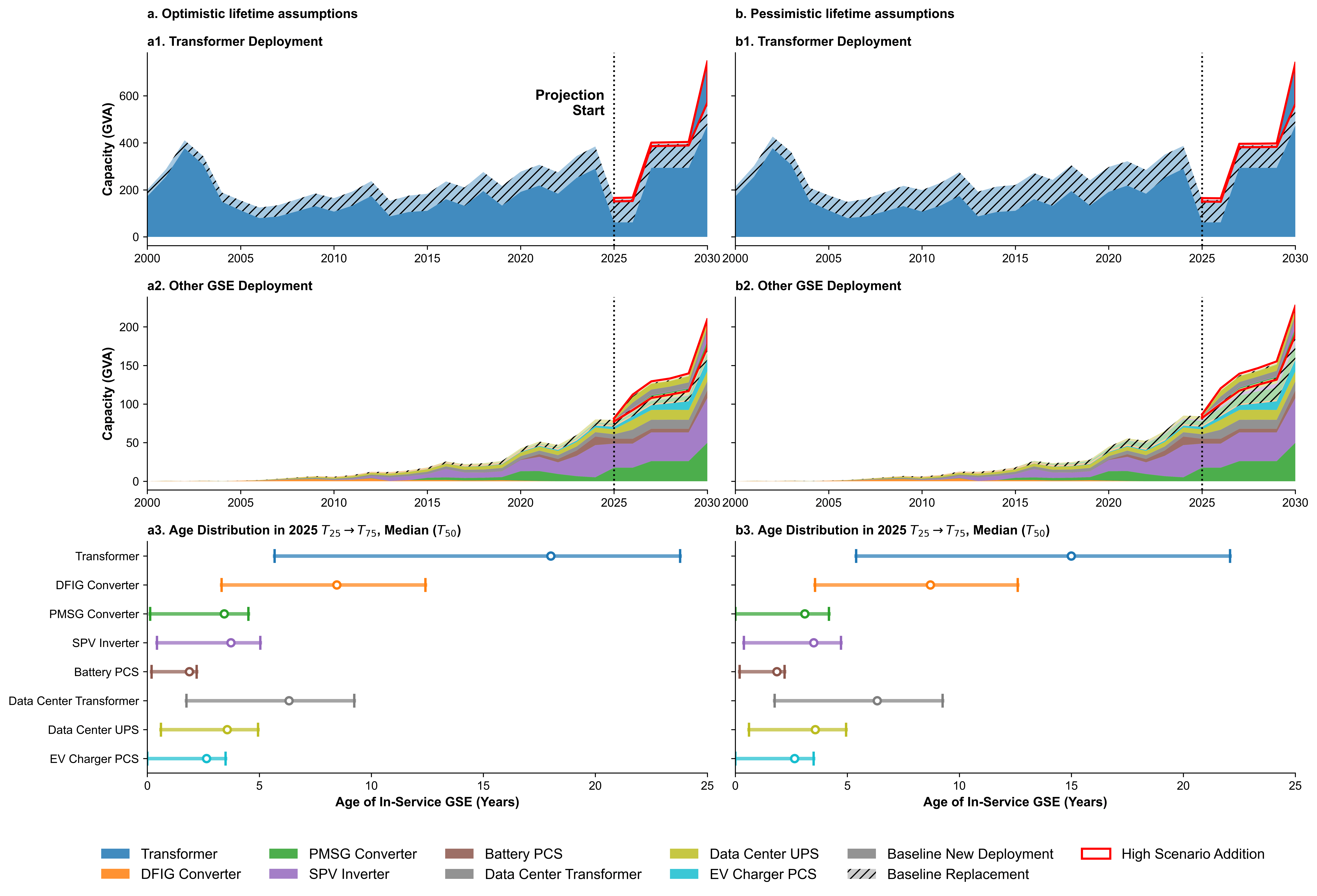}
    \caption{\textbf{Historical and projected capacity additions and age structure of GSE under (a) optimistic and (b) pessimistic lifetime assumptions.} (a1,b1) Transformer capacity additions from 2000 to 2030, separated into new deployment and replacement demand. (a2,b2) Capacity additions for other GSE classes. (a3,b3) Age distribution of in-service GSE in 2025, shown by the 25th percentile, mean, and 75th percentile.}
    \label{fig:2}
\end{figure}

We organize the Results from GSE demand until 2030 formation to material exposure and system-level implications. We first reconstruct historical stocks, characterize lifetime heterogeneity, and project future GSE demand from both replacement and new deployment. We next quantify the BOM and material intensities of GSE, and then trace how these requirements are allocated across economic sectors, identifying geographic concentration and cross-sectoral dependencies. We subsequently assess GSE shortages and material bottlenecks under different load-growth scenarios with optimistic and pessimistic lifetime assumptions, and test the robustness of these findings under alternative mitigation measures, including grid-enhancing technologies (GETs) and trade disruptions.

\subsection*{Historical stock reconstruction, lifetime heterogeneity, and future demand for GSE}
\label{sec:2.1}

We first analyze the historical stock of GSE, how it has accumulated across supply-side and load-side applications, and how future demand is shaped by both new deployment and replacement. Using observed generation  \cite{eia860_data} and projected future generation resource development \cite{osti_2496240, epri_powering_intelligence_2026, wood20232030} and mappings from these resources to their associated GSE requirements, we reconstruct the evolution of in-service GSE stocks from the start of the historical record. We then use lifetime-based survival functions (Fig.~\ref{fig:2_1} a), calibrated to reported GSE lifetime and reliability estimates, to project future new deployment and replacement needs with optimistic (i.e., empirical average lifespans) and pessimistic (i.e., shortened lifespans due to severe operating) lifetime assumptions (Fig.~\ref{fig:2_1} b).

The results show that future GSE demand is driven by two distinct but overlapping dynamics: aging-driven replacement of long-lived conventional assets and rapid expansion of newer power-electronic equipment. With optimistic lifetime assumption, transformer additions dominate total interfacing capacity throughout the study period (Fig.~\ref{fig:2} a1), reflecting both the large  legacy base and continued growth in generation, electrification, and data centers. An increasing share of these additions by 2030 is attributable to replacement, revealing that stock turnover becomes an increasingly important driver of future transformer demand and may affect the GSE availability for new deployments.

By contrast, non-transformer GSE remains limited in historical capacity but grows rapidly after 2015 (Fig.~\ref{fig:2} a2). The fastest increases occur in solar PV inverters, wind converters, battery PCS, EV charger PCS, and data center UPS, consistent with the expansion of inverter-based resources, storage, transport electrification, and digital infrastructure. Although the aggregate deployment of these GSE remains smaller than that of transformers by 2030, their growth rates are substantially steeper.

The 2025 in-service fleet also exhibits pronounced heterogeneity in age structure (Fig.~\ref{fig:2} a3). Transformers are the oldest assets on average and span a broad age range, reflecting long service lives and accumulated historical buildout. By contrast, battery PCS, EV charger PCS, solar photovoltaic inverters, and data center transformers are substantially younger, indicating more recent deployment and a still-maturing installed base. Wind converter fleets occupy an intermediate but wider age range, suggesting more mixed service ages.

These differences in age structure imply different sources of future demand. Long-lived transformer assets are increasingly driven by the replacement of the existing fleet, whereas younger and rapidly scaling power-electronic (PV inverters, wind converters, battery PCS, data center UPS, and EV charger PCS) assets are driven primarily by expansion, with replacement pressures emerging on shorter timescales. The pessimistic lifetime case preserves the same overall pattern while shifting more demand toward faster replacement within the first 10–15 years after deployment (Fig.~\ref{fig:2} b2).

\subsection*{Material intensity of GSE}
\label{sec:2.2}

\begin{figure}[t!]
    \centering
    \includegraphics[width=\linewidth]{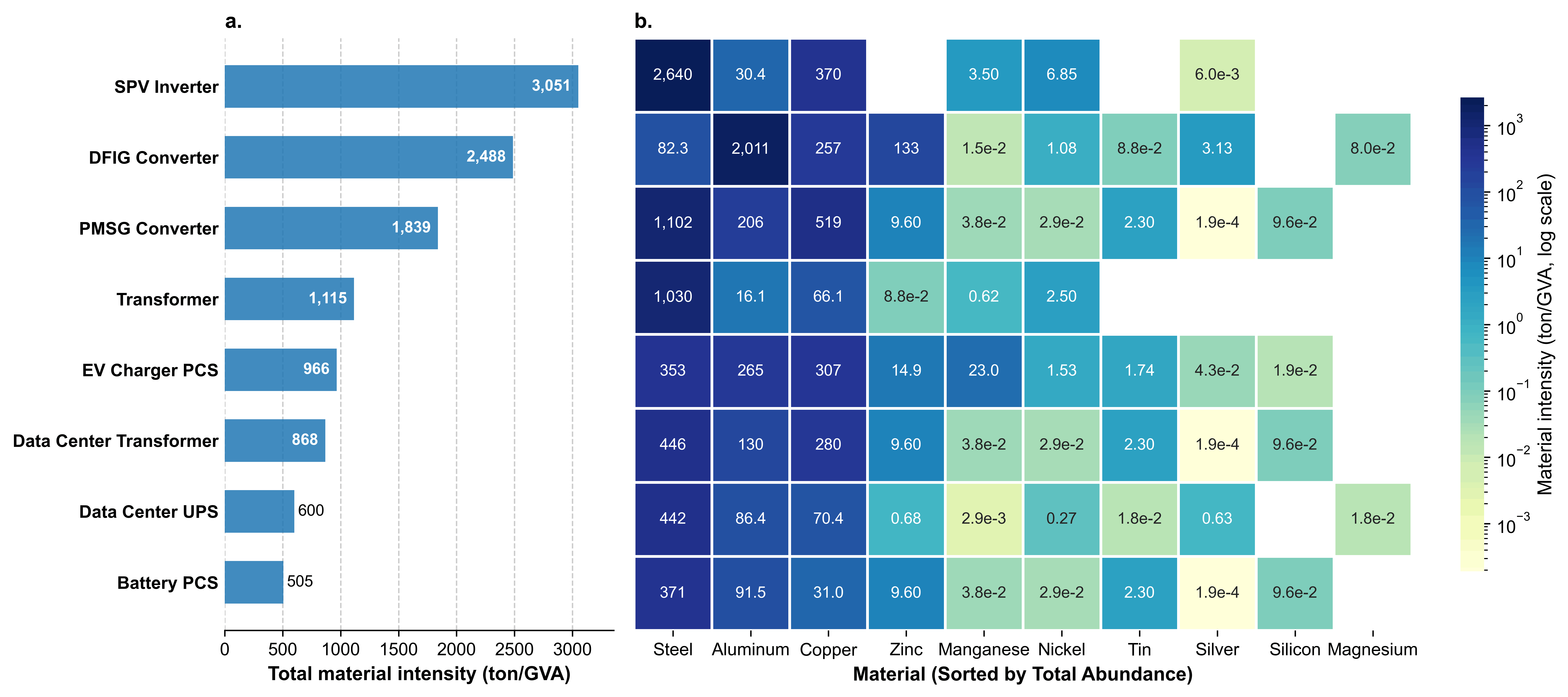}
    \caption{\textbf{Material intensity of major GSE.}
    (a) Total material intensity per unit of capacity for each equipment type, reported in ton/GVA. (b) Material composition of each technology, with cell values showing material intensity by equipment type on a logarithmic scale.}
    \label{fig:4}
\end{figure}

We next analyze how different classes of GSE translate installed electrical capacity into critical material requirements. GSE demand and shortages are reported in GVA because they represent required and unmet equipment capacity, whereas physical material requirements are reported in ton. Using the compiled equipment-level BOM data, we compare the resulting material-intensity coefficients for major supply- and load-side GSE classes on a common ton/GVA basis. Further details on material selection, source harmonization, BOM compilation, and capacity normalization are provided in the Methods section.

Figure~\ref{fig:4} summarizes the material intensity of major GSE classes and highlights substantial variation in both total material intensity and composition across equipment types. Across most equipment types, steel and copper account for the largest share of embodied mass, while the contribution of aluminum and lower-mass specialty inputs, such as zinc, tin, nickel, manganese, silver, silicon, and magnesium, depends strongly on equipment architecture. Conventional transformers are strongly steel-dominant, with 1030 ton/GVA of steel and 66.1 ton/GVA of copper, while aluminum remains comparatively limited at 16.07 ton/GVA. Data center transformers exhibit a similar composition but with substantially higher copper and aluminum intensity, reaching 445.728 ton/GVA of steel, 280.416 ton/GVA of copper, and 129.792 ton/GVA of aluminum. This indicates that load-side transformers (e.g., for data centers) also contribute to upstream material demand. 

Power-electronic equipment exhibits a distinct pattern. Solar PV inverters are the most steel-intensive assets among the GSE classes considered, with 2640 ton/GVA of steel, 370 ton/GVA of copper, and 30.4 ton/GVA of aluminum. Wind converter architectures are also materially heterogeneous. Permanent magnet synchronous generator (PMSG) converters contain 1102 ton/GVA of steel and 519.1 ton/GVA of copper, making them the most copper-intensive asset among the GSE classes considered, whereas doubly fed induction generator (DFIG) converters are dominated by aluminum at 2011 ton/GVA, together with 256.6 ton/GVA of copper. These differences show that upstream material exposure can vary substantially even within the same generation technology.

Battery and load-side conditioning systems add a further layer of demand. Battery PCS contains 370.92 ton/GVA of steel, 91.476 ton/GVA of aluminum, and 30.954 ton/GVA of copper. EV charger PCS is comparatively balanced across major bulk materials, with 353.16 ton/GVA of steel, 306.508 ton/GVA of copper, and 264.979 ton/GVA of aluminum, indicating that transport electrification can create broad-based material demand rather than dependence on a single dominant input. Data center UPS similarly combines high steel intensity, at 441.6 ton/GVA, with nontrivial aluminum and copper requirements of 86.4 and 70.4 ton/GVA, respectively.

Although several materials contribute less to total mass than steel, aluminum, and copper, they remain important for supply chain exposure because they are concentrated in specific equipment classes and often serve specialized and non-substitutable electrical, thermal, or electrochemical functions. For example, SPV inverters contain 6.85 ton/GVA of nickel, DFIG converters contain 133.4 ton/GVA of zinc and 3.132 ton/GVA of silver, and EV charger PCS contains 22.999 ton/GVA of manganese. Data center transformers and battery PCS also contain measurable tin, silicon, and zinc. These materials do not dominate the total tonnage, but they broaden the range of upstream sectors required to support GSE deployment and may become relevant next-tier constraints when demand for bulk materials is already high.

\subsection*{Material allocation and sourcing for GSE manufacturing}
\label{sec:2.3}

\begin{figure}[t!]
    \centering
    \includegraphics[width=\linewidth]{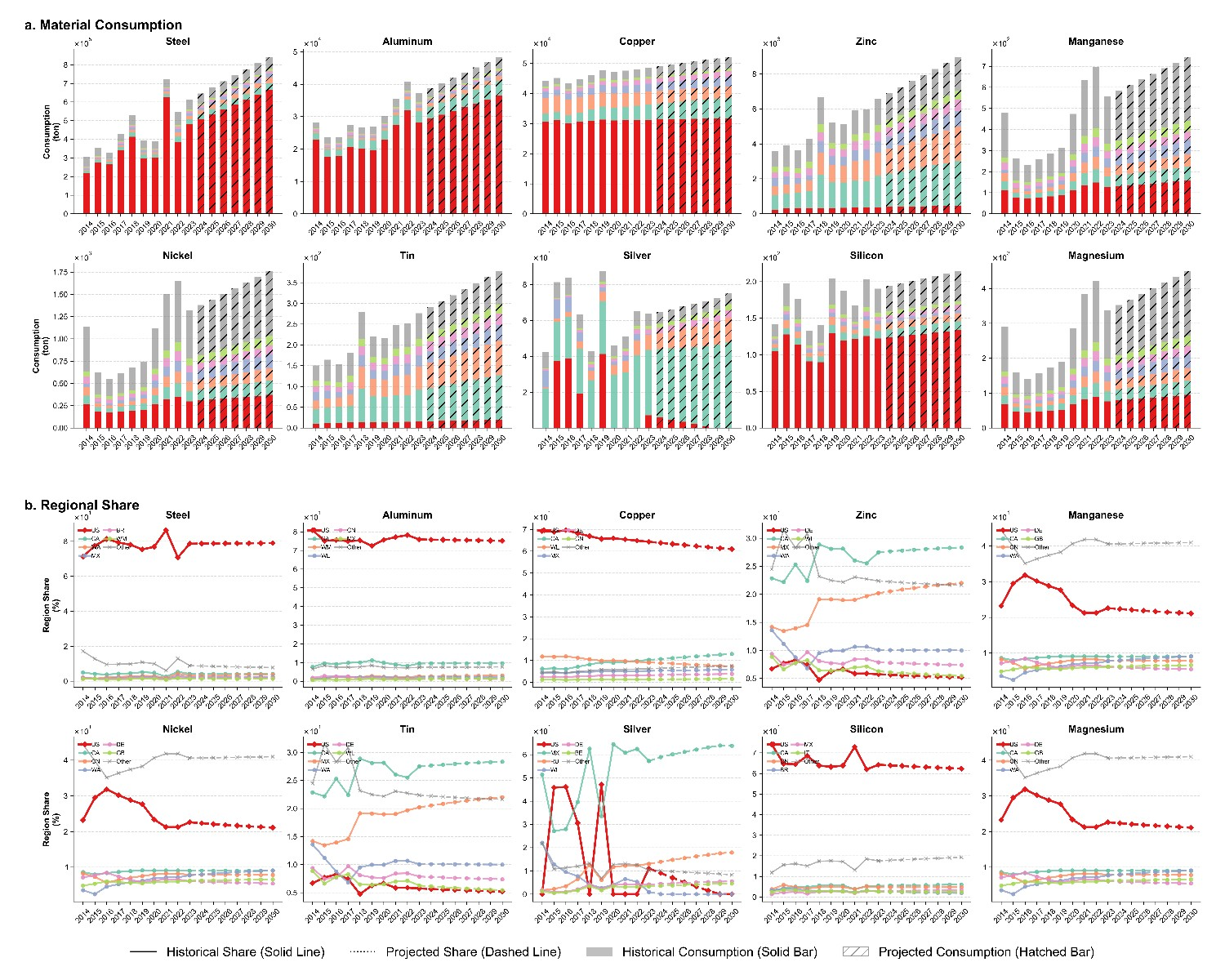}
    \caption{\textbf{Historical and projected GSE-relevant material flows and regional supply shares.}
(a) Material consumption associated with GSE manufacturing, shown as solid bars for historical values and hatched bars for projected values. (b) Regional supply shares for the same materials, shown as solid lines for historical values and dashed lines for projected values. Colors denote major supplying regions, and the remaining share is grouped as ``Other.''}
    \label{fig:5}
\end{figure}

We then analyze the upstream production and regional sourcing structure associated with the materials used in U.S.-based GSE manufacturing. The BOM-derived material intensity coefficients determine the physical material requirements of GSE deployment, while a multi-regional\footnote{Here, ``region'' denotes the country-level and aggregated rest-of-world units represented in EXIOBASE, a global multiregional supply--use and input--output database used to trace production, trade, and embodied resource flows across countries and sectors \cite{stadler_2025_15689391}.} supply-use accounting framework is adopted to independently characterize the corresponding upstream production and sourcing structure and estimate GSE manufacturing material availability. We then trace regional sourcing shares to reveal the geographic concentration of these dependencies and their exposure to disruption \cite{stadler2018exiobase, stadler2021exiobase}. Detailed allocation, scaling, and extrapolation procedures are provided in the Methods section.

Figure~\ref{fig:5} shows two consistent patterns across the  materials considered. First, total material consumption associated with GSE manufacturing increases over time for nearly all materials (Figure~\ref{fig:5} a). Second, the regional composition of supply differs markedly across bulk and next-tier materials (Figure~\ref{fig:5} b).

As shown in Figure~\ref{fig:5} a, bulk materials dominate total embodied GSE consumption. After applying the GSE-specific allocation, steel remains by far the largest material flow, rising from 486 to 564 thousand tons between 2024 and 2030. Over the same period, aluminum rises from 35.4 to 41.4 thousand tons, copper from 8.28 to 9.20 thousand tons, and zinc from 1.93 to 2.48 thousand tons. Although smaller in absolute scale, nickel, magnesium, silicon, manganese, tin, and silver all exhibit clear upward trends through 2030, indicating that rapid GSE deployment expands not only bulk material demand but also dependence on a broader set of specialized upstream material inputs.

Figure~\ref{fig:5} b shows that sourcing patterns are not uniform across materials. Steel and aluminum remain strongly concentrated in the U.S. throughout both the historical and projected periods, with the U.S. share near 79\% for steel and 75\% for aluminum by 2030. Copper is also U.S.-centered, but less concentrated, with the U.S. share declining from about 70\% in the mid-2010s to roughly 61\% by 2030, alongside rising contributions from Canada, Mexico, and other supplying regions. This implies that even for the three core bulk materials, future GSE manufacturing remains partly dependent on foreign supply.

Several next-tier materials exhibit more diversified or import-dependent sourcing structures. Zinc is primarily sourced from Canada and Mexico with 2030 shares of about 28\% and 22\% (the U.S. shares near 5\%). Tin follows a similar pattern. Silver is the most concentrated non-bulk material, with Mexico supplying nearly 64\% by 2030 and Russia contributing a further 18\%. Manganese, nickel, and magnesium each show relatively low U.S. shares of about 21\% in 2030 and rely on a broader set of external suppliers, including Canada, Western Asia, China, Australia, and the United Kingdom. Silicon stands apart from this trend, where over 60\% remains in the U.S.

Taken together, these results indicate that material exposure in GSE manufacturing arises through two different channels. One is the growing scale of demand for bulk materials, especially steel, aluminum, and copper, as deployment expands over time. The other is the reliance on regionally concentrated or import-dependent supply for several smaller-volume materials, which may increase vulnerability to trade disruption even when total tonnage is modest. As a result, future GSE deployment depends not only on aggregate material availability, but also on the structure and resilience of the supply chains that provide those materials.

\subsection*{GSE shortages and material bottlenecks under baseline and high-growth scenario with optimistic lifetime assumption}
\label{sec:2.4}

To determine how material availability translates into feasible GSE deployment, we first use the baseline and high-growth scenarios to specify annual GSE requirements from projected generation and load growth together with equipment replacement; these requirements are treated as target demand. The GSE deployment model then allocates available materials across equipment classes and determines how much of each requirement can be supplied, with any remaining requirement reported as unmet GSE demand. When material supply is insufficient to meet all the requirements, the model first prioritizes proportionally complete deployment across GSE classes and then allocates the remaining feasible deployment according to an equipment hierarchy, with grid-enabling transformers prioritized over generation-side GSE, followed by load-side GSE. This structure is intended to preserve the supporting grid infrastructure needed for broader system deployment rather than allowing scarce materials to be concentrated in a single equipment class. These scenarios differ along three demand drivers: overall electricity demand and associated generation expansion, data center build-out, and EV adoption. The baseline scenario assumes moderate growth across all three components, whereas the high-growth scenario assumes faster load growth, more extensive data center build-out by 2030, and a larger EV fleet. Additional details on the deployment hierarchy and model formulation are provided in the Methods section.

\begin{figure}[t!]
    \centering
    \includegraphics[width=\linewidth]{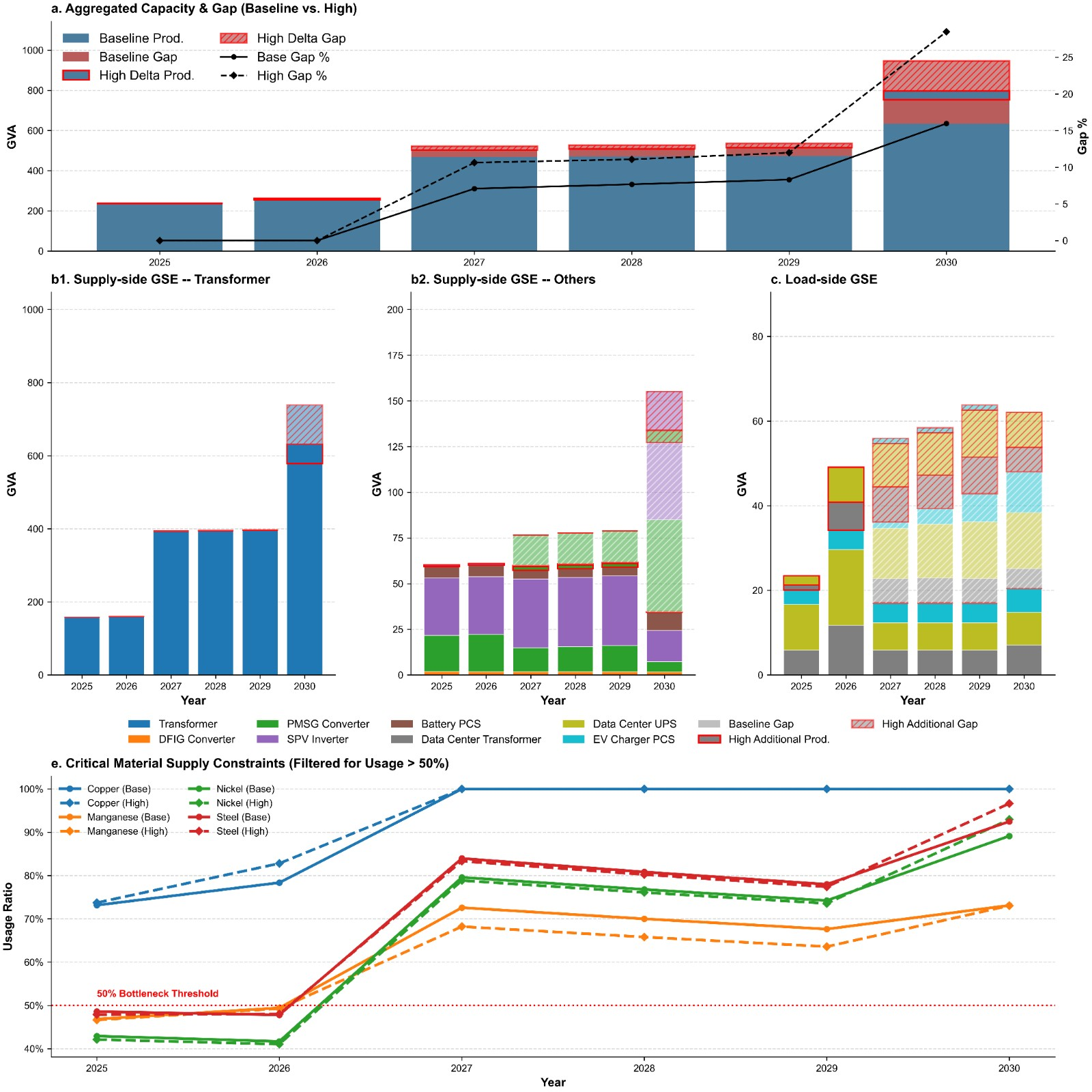}
    \caption{\textbf{Baseline and high-growth scenario comparisons of GSE deployment, unmet demand gaps, and material usage with optimistic lifetimes.}
(a) Aggregate GSE deployment, unmet demand, and gap ratios. (b1) Supply-side GSE results for transformers. (b2) Supply-side GSE results for other equipment categories. (c) Load-side GSE results. (d) Annual usage ratios of selected critical materials, together with the 50\% bottleneck threshold.}
    \label{fig:6}
\end{figure}

Figure~\ref{fig:6} shows three main findings. First, aggregate GSE shortages do not appear in 2025 or 2026, but emerge from 2027 onward and widen substantially by 2030. Second, shortages initially concentrate in non-transformer supply-side electronics and load-side GSE, while transformers remain largely satisfied until the end of the horizon. Third, these equipment shortfalls coincide with increasingly tight upstream material usage, with copper becoming fully binding and steel and nickel approaching saturation.

At the aggregate level (Figure~\ref{fig:6} a), total GSE demand in the baseline scenario rises from 237.6 GVA in 2025 to 254.1 GVA in 2026, then increases to 503.5 GVA in 2027 and 753.6 GVA by 2030, reflecting the combined effect of new deployment and replacement demand. No unmet demand appears in 2025 or 2026. This margin disappears in 2027, when unmet demand reaches 35.7 GVA, equivalent to 7.1\% of total requirements. This gap  gradually grows to 42.9 GVA (8.3\%) in 2029, before expanding sharply to 120.3 GVA (16.0\%) in 2030. The high-growth scenario follows the same qualitative pattern but deteriorates more rapidly, where the total required volume rises to 946.6 GVA by 2030 and the aggregate shortfall reaches 269.6 GVA (28.5\%) in 2030. Relative to the baseline scenario, high-growth increases 2030 total GSE demand by 25.6\% and more than doubles unmet deployment capacity.

Disaggregated results (Figure~\ref{fig:6} b1-c) show that shortages emerge first outside the transformer category. In the baseline scenario, transformer demand remains fully met through 2029, while deficits from 2027 to 2029 are concentrated in a narrower set of other GSE categories. On the supply side, near-term shortages are driven primarily by PMSG converters, with unmet demand reaching 17.1 GVA in 2029 (54.4\%). By contrast, SPV inverters, battery PCS, and replacement DFIG converters remain fully served during this period. On the load side, shortages appear in data center transformers, data center UPS, and EV charger PCS. In 2027, unmet demand reaches 5.8 GVA (49.5\%) for data center transformers, 11.9 GVA (64.8\%) for data center UPS, and 1.5 GVA (24.3\%) for EV charger PCS. These deficits persist through 2029. By 2030, the baseline shortfall remains dominated by non-transformer technologies: unmet demand reaches 50.5 GVA (89.9\%) for PMSG converters and 42.3 GVA (71.4\%) for SPV inverters on the supply side, while load-side shortages rise to 4.7 GVA (40.2\%) for data center transformers, 13.2 GVA (63.0\%) for data center UPS, and 9.7 GVA (63.1\%) for EV charger PCS.

The high-growth scenario amplifies downstream shortages and eventually extends them to transformers. Through 2029, transformer demand is still fully met, but load-side GSE shortages are substantially larger than in the baseline scenario, reaching 46.8 GVA (78.8\%) in 2029. Over the same period, unmet demand in other supply-side electronics remains persistent and reaches 16.5 GVA (21.4\%) in 2029. The key shift occurs in 2030, when transformer demand itself exceeds feasible production. Under the high-growth scenario, transformer production reaches 631.5 GVA, but an additional 107.3 GVA remains unmet, corresponding to a 14.5\% transformer-specific gap. By 2030, shortages also intensify in both other equipment groups, with unmet demand reaching 41.6 GVA (67.1\%) for load-side GSE and 120.6 GVA (82.8\%) for other supply-side electronics. This marks a transition from shortages concentrated in peripheral or fast-scaling equipment to shortages affecting the core grid-enabling asset class itself.

The material usage results (Figure~\ref{fig:6} d) mirror this progression. In the baseline scenario, copper usage is already high in the near term, rising from 73.2\% in 2025 to 78.4\% in 2026, and reaching 100\% from 2027 onward, indicating a binding copper constraint over the entire late-horizon period. Steel usage increases from 48.6\% in 2025 to 84.0\% in 2027 and remains elevated thereafter, reaching 92.5\% in 2030. Nickel follows a similar trajectory, rising from 42.3\% in 2025 to 89.1\% in 2030. Manganese also exceeds the 50\% threshold in 2027 and remains between 68\% and 73\% through 2030. The high-growth scenario preserves the same ordering of constraints but pushes several materials closer to exhaustion. By 2030, copper remains fully binding at 100\%, while steel rises to 96.7\% and nickel to 93.0\%, both above the corresponding baseline levels. Manganese remains elevated at 73.1\%. Together, these results indicate that the widening GSE shortfall is associated first with copper scarcity and then with increasingly tight competition for steel- and nickel-intensive supply chains.

The sequencing of shortages reflects the interaction of demand growth, material requirements, and the model's allocation structure. First, rapidly expanding non-transformer GSE, particularly solar PV inverters, PMSG converters, and load-side equipment, creates substantial new deployment requirements. Second, these equipment classes compete for common upstream materials as copper supply becomes exhausted and steel and nickel become increasingly constrained. Third, once proportionally complete deployment across GSE classes can no longer be maintained, the allocation hierarchy prioritizes grid-enabling transformers over generation-side and load-side GSE. Early shortages therefore appear primarily in non-transformer equipment, while transformer shortages arise only when overall material pressure becomes sufficiently large that even the prioritized transformer requirement cannot be fully met. The later appearance of transformer shortages should therefore be interpreted partly as a consequence of this allocation structure rather than as evidence that transformer supply is intrinsically less constrained.

\subsection*{GSE shortages and material bottlenecks under baseline and high-growth scenario with pessimistic lifetime assumption}
\label{sec:2.5}

\begin{figure}[t!]
    \centering
    \includegraphics[width=\linewidth]{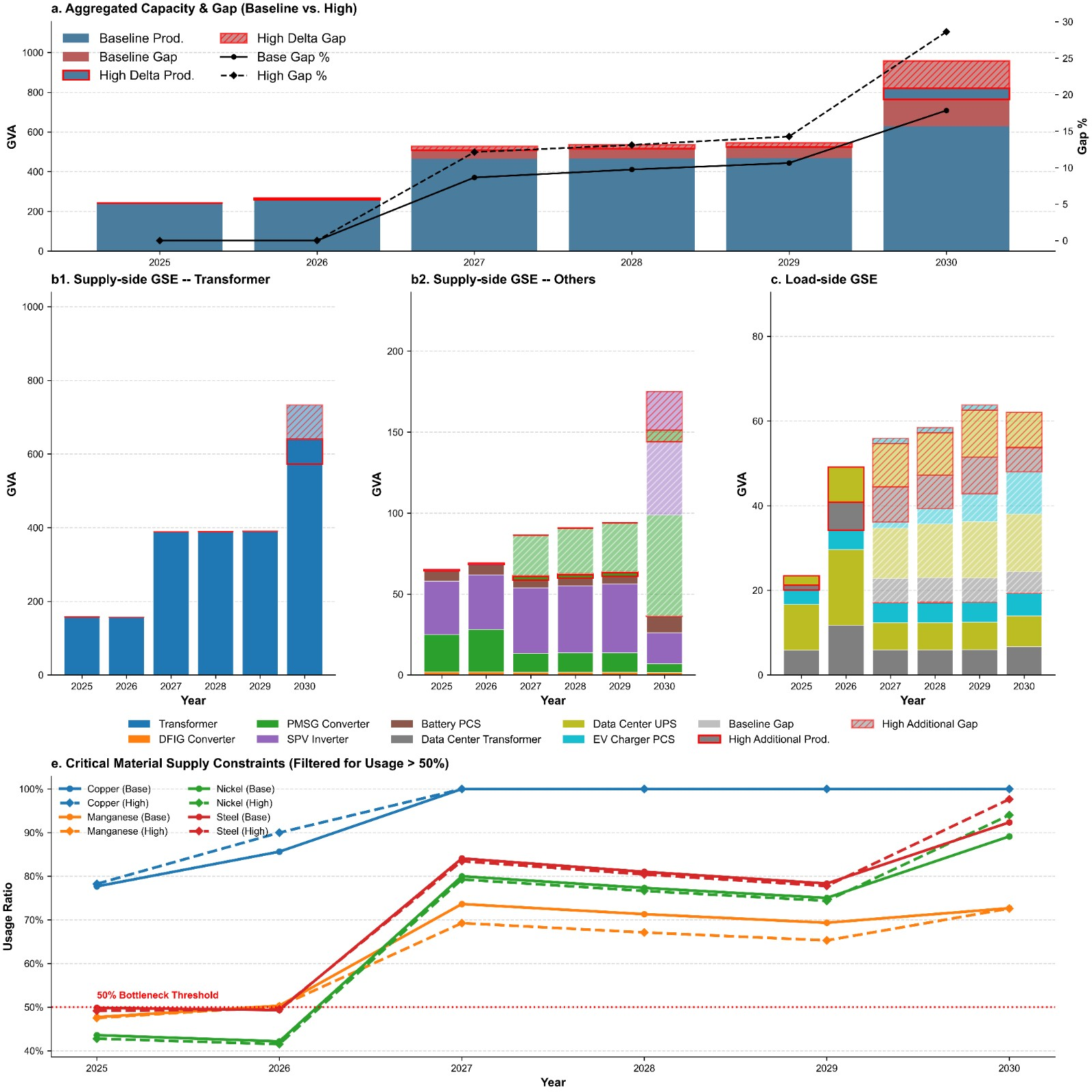}
    \caption{\textbf{Baseline and high-growth scenario comparisons of GSE deployment, unmet demand gaps, and material usage with pessimistic lifetimes.}
(a) Aggregate GSE deployment, unmet demand, and gap ratios. (b1) Supply-side GSE results for transformers. (b2) Supply-side GSE results for other equipment categories. (c) Load-side GSE results. (d) Annual usage ratios of selected critical materials, together with the 50\% bottleneck threshold.}
    \label{fig:7}
\end{figure}

We re-evaluate the baseline and high-growth scenarios with pessimistic lifetime assumptions while keeping other factors unchanged to isolate the effect of earlier GSE retirements and replacements on the timing and severity of shortages. Figure~\ref{fig:7} shows that shorter lifetimes do not change which GSE classes become constrained first, but shift more replacement demand into the 2025--2030 horizon and amplify the shortages identified under optimistic lifetimes. This effect is strongest for non-transformer GSE, especially technologies with younger and faster-growing installed bases, while the transformer bottleneck changes only modestly.

At the aggregate level (Figure~\ref{fig:7} a), the pessimistic lifetime assumption increases unmet demand in both baseline and high-growth scenarios. In the baseline scenario, unmet demand rises from 35.7 GVA (7.1\%) to 44.0 GVA (8.7\%) in 2027 and from 120.3 GVA (16.0\%) to 136.4 GVA (17.8\%) in 2030. In the high-growth scenario, the corresponding shortfall rises from 54.7 GVA (10.6\%) to 63.0 GVA (12.1\%) in 2027 and from 269.6 GVA (28.5\%) to 274.1 GVA (28.6\%) in 2030. These differences indicate that shortening equipment lifetimes primarily advances and enlarges unmet demand without changing the overall shortage sequence.

This amplification is concentrated in replacement-sensitive non-transformer GSE. With pessimistic lifetimes, shortages increase most clearly for PMSG converters and SPV inverters, with smaller but persistent increases in load-side GSE such as data center transformers, data center UPS, and EV charger PCS. By contrast, the transformer bottleneck is not further amplified within the modeled time period. This reflects a timing shift in replacement demand rather than any structural easing of transformer constraints, as shorter assumed lifetimes move a larger share of replacement needs before 2025. In the high-growth scenario, for example, the 2030 transformer shortfall is 92.7 GVA (12.6\%) with pessimistic lifetimes, compared with 107.3 GVA (14.5\%) with optimistic lifetimes.

Furthermore, the upstream material ranking remains unchanged. Copper becomes fully binding from 2027 onward, with steel and nickel emerging as the next-tightest constraints under both baseline and high-growth scenarios. Ultimately, the pessimistic lifetime scenario reinforces the primary conclusion drawn from the optimistic lifetimes: the accelerated, replacement-driven demand for rapidly scaling non-transformer GSE creates a critical supply bottleneck.

\subsection*{Sensitivity of material bottlenecks to material availability assumptions}
\label{sec:sensitivity_material}

To examine how uncertainty in upstream material availability affects the identified bottlenecks, we repeat the reference high-growth scenario with optimistic lifetime assumptions using the historical lower and upper allocation bounds defined in the Methods. These bounds represent the historical variation in the share of the broader electrical-machinery sector attributed to GSE. We also test greater copper availability because copper is the first material to become fully utilized in our analysis (Figs.~\ref{fig:6}--\ref{fig:7}) and therefore determines the onset of the material constraint in the reference scenario. Given the near-term horizon of this study, we focus this sensitivity on material availability rather than changes in equipment design, technology innovation or material intensity. Specifically, we increase the copper supply accessible to GSE by 6.75\% while holding the availability of other materials unchanged, providing a targeted sensitivity on the material that first becomes binding in the reference scenario. This value is informed by the IEA Announced Pledges Scenario, in which clean-energy technologies account for approximately 24.4\% of global copper demand in 2023 and 38.6\% in 2030, corresponding to an annualized increase of approximately 6.75\% in their share of copper demand \cite{iea2024copper}. We use this change as a proxy for the extent to which copper could become more accessible to GSE as material demand shifts toward rapidly expanding power grids. Accordingly, the sensitivity represents a change in the sectoral allocation of available copper rather than an equivalent increase in global copper production. Together with the historical allocation bounds, these cases examine how uncertainty in upstream material allocation and greater access to the initially binding material affect both the magnitude of GSE shortages and the sequence of material constraints.

\begin{table*}[t!]
\centering
\caption{\textbf{Sensitivity of GSE shortages and material bottlenecks to alternative supply assumptions in 2030.}}
\label{tab:material_sensitivity}
\small
\setlength{\tabcolsep}{2pt}
\begin{tabular}{lcccc}
\toprule
\textbf{Sensitivity case}
& \textbf{Unmet GSE demand}
& \textbf{First binding}
& \textbf{Limiting material}
& \textbf{2030 usage ratio} \\
&
\textbf{[GVA (\%)]}
& \textbf{material (year)}
& \textbf{in 2030}
& \textbf{Cu / steel / Ni (\%)} \\
\midrule
Reference
& 269.6 (28.5)
& Cu (2027)
& Cu
& 100.0 / 96.7 / 93.0 \\

Lower bound
& 347.2 (36.7)
& Cu (2027)
& Cu
& 100.0 / 94.7 / 90.9 \\

Upper bound
& 202.3 (21.4)
& Cu (2027)
& Cu
& 100.0 / 98.1 / 94.5 \\

Cu response
& 246.5 (26.0)
& Cu (2027)
& Steel
& 96.4 / 100.0 / 96.3 \\

Lower bound + Cu response
& 314.1 (33.2)
& Cu (2027)
& Steel
& 98.0 / 100.0 / 96.1 \\

Upper bound + Cu response
& 187.8 (19.8)
& Cu (2027)
& Steel
& 95.3 / 100.0 / 96.4 \\
\bottomrule
\end{tabular}
\\[2pt]
\begin{minipage}{\linewidth}
\footnotesize
\textbf{Note:} The reference case corresponds to the high-growth scenario with optimistic lifetime assumptions reported in Fig.~6. All sensitivity cases use the same demand and lifetime assumptions. The lower and upper bounds correspond to the minimum and maximum historical shares of GSE material allocation within the broader electrical-machinery sector over 2012--2021, as defined in the Methods. ``Cu response'' increases GSE-accessible copper availability by 6.75\% while holding the availability of other materials unchanged.
\end{minipage}
\end{table*}

The historical allocation bounds primarily affect the magnitude of the material constraints, rather than their order. Relative to the reference scenario, the lower allocation bound increases unmet GSE demand in 2030 from 269.6~GVA (28.5\%) to 347.2~GVA (36.7\%), whereas the upper allocation bound reduces it to 202.3~GVA (21.4\%). The ordering at the onset of the constraint is unchanged: copper is the first material to bind in 2027 under both allocation bounds and remains the limiting material in 2030 when its availability is otherwise unchanged. This indicates that uncertainty in the upstream allocation affects the severity of the resulting GSE shortage more strongly than the identity or timing of the initial material constraint.

Increasing the copper supply accessible to GSE by 6.75\% reduces the 2030 shortfall from 269.6 to 246.5~GVA under the reference allocation. Copper still becomes constrained first in 2027, indicating that greater copper availability delays and moderates its effect rather than removing it as the near-term constraint. As more GSE can subsequently be deployed, demand for the other materials embodied in the same equipment also increases. By 2030, steel therefore reaches full utilization, while copper and nickel remain close to their limits at 96.4\% and 96.3\%, respectively. The same pattern appears under both historical allocation bounds, with steel becoming limiting by 2030 when additional copper is available.

The sensitivity results therefore reveal an evolving sequence of material constraints. Copper consistently governs the onset of scarcity across all cases examined, reinforcing its importance for near-term GSE deployment. Greater access to copper allows additional deployment but, in turn, brings steel and nickel closer to their own supply limits, with steel becoming the limiting material by the end of the study period. The broader implication is that alleviating the initial copper constraint can reduce GSE shortages, while further expansion increasingly depends on the joint availability of several materials.

\subsection*{Sensitivity to trade disruption and dynamic transformer rating (DTR)}
\label{sec:2.6}

Building on the high-growth scenario with optimistic lifetime assumption, we evaluate two policy-relevant sensitivities that act on distinct parts of the GSE constraint. In the trade-disruption scenario, material supply from geopolitically less predictable partner regions \footnote{Here, ``geopolitically less predictable partner regions'' refers to regions outside of the North Atlantic Treaty Organization (NATO), Major Non-NATO Allies (MNNA), the American Nations Energy Alliance (AMNEA), the North American Free Trade Agreement (NAFTA), and permanently neutral states such as Switzerland and Austria.} is reduced by 70\% \cite{bolhuis2023fragmentation}. In the DTR scenario, required deployment of transformers and data center transformers is reduced by 10\%, reflecting the ability of GETs like DTR to unlock additional utilization from existing transformer capacity \cite{daminov2021assessment}.

The two sensitivities produce a strongly asymmetric response. Trade disruption substantially worsens the shortages, whereas DTR provides only partial relief. Relative to the high-growth optimistic reference scenario in 2030, trade disruption increases total unmet GSE demand by 73.0 GVA, raising the aggregate shortfall from 269.6 GVA (28.5\%) to 342.6 GVA (36.2\%). By contrast, DTR reduces total unmet demand by 75.6 GVA, lowering the aggregate shortfall to 193.9 GVA (22.3\%). Although these aggregate changes are similar in magnitude, their structure differs. 

The trade-disruption scenario primarily amplifies the transformer bottleneck. In 2030, transformer unmet demand rises by 77.4 GVA, increasing from 107.3 GVA (14.5\%) to 184.7 GVA (25.0\%). By contrast, unmet demand for all other GSE changes little in aggregate, decreasing slightly by 4.3 GVA because tighter material access shifts more of the shortage burden toward transformers within the model's deployment hierarchy. The associated material response is also clear. Copper remains fully binding, while nickel usage rises by 7.0 percentage points to full saturation in 2030 and manganese usage rises by 8.2 percentage points to 81.3\%. Thus, trade disruption does not alter the ordering of constraints, but compresses material availability enough to make the existing shortages substantially more severe.

\begin{table}[t!]
\centering
\caption{\textbf{Sensitivity of GSE shortages and material usage to trade disruption and DTR in 2030, relative to the high-growth optimistic reference scenario.}}
\label{tab:sensitivity_delta_2030}
\small
\setlength{\tabcolsep}{5pt}
\begin{tabular}{lcc}
\toprule
\textbf{Metric} & \textbf{Trade disruption} & \textbf{DTR} \\
\midrule
Change in total unmet GSE demand (GVA) & +73.0 & -75.6 \\
Change in total unmet GSE demand (\%) & +7.7 & -6.2 \\
Change in transformer unmet demand (GVA) & +77.4 & -73.9 \\
Change in transformer unmet demand (\%) & +10.5 & -9.5 \\
Change in all other GSE unmet demand (GVA) & -4.3 & -1.8 \\
Change in all other GSE unmet demand (\%) & -2.1 & -0.2 \\
Change in nickel usage ratio (\%) & +7.0 & 0.0 \\
Change in manganese usage ratio (\%) & +8.2 & 0.0 \\
\bottomrule
\end{tabular}
\\[2pt]
\begin{minipage}{\linewidth}
\footnotesize
\textbf{Note:} Values are changes relative to the high-growth optimistic scenario in 2030. Percentage rows report percentage-point changes. ``All other GSE'' includes all non-transformer categories.
\end{minipage}
\end{table}

DTR acts in the opposite direction, but only on one layer of the bottleneck stack. In 2030, transformer unmet demand falls by 73.9 GVA, from 107.3 GVA (14.5\%) to 33.4 GVA (5.0\%), and the data center transformer gap also declines. However, non-transformer shortages remain nearly unchanged, with unmet demand for all other GSE falling by only 1.8 GVA in aggregate. The main late-horizon deficits in PMSG converters, SPV inverters, data center UPS, and EV charger PCS therefore persist. Material usage changes little as well: copper remains fully binding from 2027 onward, while nickel and manganese usage are essentially unchanged in 2030. These results indicate that DTR is effective as a targeted mitigation for transformer scarcity, but does not address the dominant shortages in power electronics or the underlying bulk-material constraint.

Taken together, these sensitivities show that 2030 shortages are more vulnerable to upstream geoeconomic disruption than they are responsive to incremental efficiency gains in transformer utilization. Trade disruption worsens the existing bottleneck structure by tightening access to already constrained material supply, while DTR provides meaningful but bounded relief by acting on only one equipment class. This comparison suggests that transformer-focused operational measures can relieve part of the constraint, but system-wide resilience ultimately depends on broader material access and manufacturing capability across GSE supply chains.

\section*{DISCUSSION}
\label{sec:3}

Our results show that power system expansion is no longer defined only by the addition of generation, storage, transmission, or load. It also hinges on whether the interfacing hardware needed to connect, convert, condition, and deliver electricity can be manufactured and deployed at the required scale and speed.  Specifically, expansion depends on the availability of GSE, together with the upstream material and industrial systems required to produce it, which are treated only implicitly in the current practice. Once these factors are represented explicitly, power system expansion becomes not only a planning problem of capacity choice, but also a coupled problem of deployment timing, replacement needs, manufacturing capability, and material availability.

This perspective matters because the modern grid is increasingly shaped not only by long-lived electromagnetic assets, but also by faster-turnover, electronics-intensive equipment classes, including converters, inverters, power conversion systems, UPS systems, and load-side transformers \cite{byles2023sustainable}. These assets differ sharply in material composition, replacement timing, and manufacturing dependencies. As a result, future deployment pressure does not arise solely from net-new buildout. It also arises from the interaction between continued expansion and repeated replacement within the same planning horizon. The results therefore suggest that the energy expansion or energy transition also require an industrial transition toward a broader and more rapidly cycling equipment base.

\subsection*{From critical materials to coupled equipment bottlenecks}

The results demonstrate different materials play different roles in GSE deployment. Copper emerges as the most persistent bottleneck because it is embedded across multiple GSE classes and therefore scales with almost every layer of electrification. Steel and aluminum remain critical because they anchor the bulk material base of transformers, converters, inverters, and other support equipment. By contrast, nickel, manganese, zinc, silver, silicon, and tin are less critical in aggregate tonnage, but remain important because they are concentrated in particular GSE classes. For example, nickel is concentrated in SPV inverters, manganese in EV charger PCS, zinc and silver in DFIG converters, and silicon and tin in data center transformers, PMSG converters, and battery PCS. These materials may therefore emerge as next-tier constraints once bulk-material pressure is already high.

This distinction is important because it implies that future bottlenecks are likely to be coupled across technologies and sectors. Data centers, EV charging networks, renewable generation, storage systems, and transmission expansion are often analyzed separately. In practice, they increasingly converge upstream through overlapping dependence on copper-intensive conductors, transformer steel, power-electronic manufacturing, and specialized industrial inputs. Infrastructure vulnerability therefore reflects not only geological scarcity or geopolitical risk, but also the economic structure through which materials are transformed, allocated, and embodied in equipment production. In this sense, scarcity is partly a problem of industrial coordination.

\subsection*{Industrial resilience levers beyond generation buildout}

The results point to several resilience levers that lie outside conventional capacity-centric planning. Greater standardization and modularity may reduce vulnerability by improving interchangeability, simplifying maintenance, and enabling faster repair or refurbishment \cite{salomez2024state, pinto2025addressing}. In supply-constrained settings, equipment heterogeneity can itself become a source of fragility if it limits substitution and slows qualification of alternative supply.

Lifetime extension is a second lever, particularly for equipment classes with short lifetime intervals or rapid projected turnover. In such cases, extending usable service life can relieve pressure much like expanding manufacturing output \cite{salomez2024state}, although the relevant mechanisms differ across transformers and power-electronic systems.

The sensitivity analysis further shows that not all interventions act on the same bottleneck layer. DTR provides targeted relief for transformer scarcity, but has limited effect on the dominant shortages in non-transformer GSE or on the underlying copper constraint. Trade disruption, by contrast, directly worsens the shortages by tightening access to already constrained supply chains. This contrast suggests that system-wide resilience depends not only on operational flexibility, but also on upstream material access and manufacturing capability.

\subsection*{Implications for planning and policy}

Taken together, these findings demonstrate that GSE availability is critical  for  \textit{feasible} power system planning. Once GSE and its material dependencies are made explicit, technology choice is no longer separable from industrial feasibility. A pathway that appears cost-effective in a conventional planning model may still prove fragile if its associated interfacing hardware cannot be manufactured, replaced, or sourced on time. One practical approach is to translate planned additions of generation, storage, and large loads into the corresponding annual requirements for transformers, converters, inverters, PCS, and other GSE, while separately accounting for replacement needs from  legacy equipment. These requirements can then be compared with estimated equipment or material availability to identify years in which an otherwise feasible expansion pathway may face a deliverability constraint.

This information could be incorporated directly as annual GSE availability constraints in CEMs or used as a post-processing screen for CEM outcomes. The latter may be particularly useful where detailed manufacturing information is limited: material availability, equipment lead times, or reported procurement constraints can provide progressively more specific measures of deliverability as better data become available. The resulting assessment would distinguish between how much capacity is desirable to build and how much of the associated GSE can plausibly be supplied within the same planning period.

The timing of these constraints is also relevant. Because GSE demand reflects both new deployment and replacement of existing equipment, a portfolio that appears feasible by an endpoint year may still encounter interim equipment shortages. Where such constraints occur, planners can compare alternatives that change the timing or composition of deployment, extend the use of existing assets, coordinate procurement, or reduce equipment requirements through measures such as DTR. Our material sensitivity results further suggest that relieving one material constraint can increase the importance of others, so the availability of the GSE portfolio as a whole is more informative than tracking a single material in isolation.

These findings also suggest a role for closer coordination between electricity planning and information on equipment manufacturing and supply chains. In practical terms, this could include greater attention to domestic and allied manufacturing capability, procurement coordination, standard-setting, refurbishment pathways, and monitoring of equipment-specific material exposure alongside more familiar system metrics such as reserve margin, reliability, and cost. Such information would allow planners to identify potential GSE constraints earlier and evaluate whether alternative deployment schedules or mitigation measures can reduce their effect. More broadly, the results suggest that the pace of grid modernization will be determined not only by how quickly new generation can be built, but also by how effectively the underlying industrial base can supply the equipment required to connect and sustain it.

\subsection*{Role of circularity in GSE supply resilience}

Circularity can affect GSE supply and, by extension, the results discussed above through several distinct mechanisms. Material recycling and recovery can supplement primary material supply; refurbishment and lifetime extension can delay equipment replacement; and reuse, remanufacturing, and parts harvesting can reduce demand for newly manufactured equipment and components. These pathways therefore act on different stages of the GSE supply chain and should not be treated as a single mitigation lever.

Recycling acts primarily on upstream material availability. Metals recovered from retired equipment can provide secondary supplies of copper, aluminum, steel, and other inputs, but recoverable material does not necessarily translate into usable secondary supply; its specific contribution depends on collection and recovery rates, material quality, reverse-logistics and reprocessing capacity, and whether recovered materials can return to GSE manufacturing at sufficient scale within the 2025--2030 planning horizon \cite{nguyen2022electric,igogo2022america,kulasek2020towards}. These constraints can be particularly important for materials that require specialized processing. Transformer core steel, for example, relies on highly engineered magnetic properties and Goss texture; conventional remelting does not preserve these properties, so closed-loop recovery requires additional processing and qualification before the material can return to high-performance transformer cores \cite{elgamli2023advancements}. By contrast, some recovery pathways are more mature. Re-refining of insulating mineral oil is commercially established under existing standards \cite{rao2020some}, although it does not address the solid-material constraints emphasized here.

Refurbishment and lifetime extension reduce GSE demand by delaying equipment replacement. Their value depends on the age and condition of the installed fleet, failure rates, maintenance practices, and whether service life can be extended without compromising reliability \cite{doe2024lptresilience, roskladka2025repairable}. Their effect is greatest where replacement accounts for a substantial share of future equipment demand.

Reuse, remanufacturing, and parts harvesting can similarly reduce demand for new GSE by retaining serviceable equipment and qualified components. Unlike material recycling, these pathways preserve more of the functional value of existing assets and can reduce requirements for newly manufactured units or subassemblies \cite{de2025circular, sun2026recycling}. They may be particularly valuable when supply is constrained by specialized components, manufacturing capacity, or qualification requirements rather than by bulk material availability alone.

The present analysis does not model these circularity pathways endogenously. Secondary material supply from recycling is excluded because consistent GSE-specific estimates of collection rates, recovery yields, material quality, and the timing of recovered material returning to GSE manufacturing are not available, or scarce, across the equipment classes considered. In addition, the framework uses material availability as a proxy for GSE supply and therefore does not explicitly represent component-level reuse, remanufacturing, parts harvesting, or associated manufacturing and qualification constraints. Refurbishment and lifetime extension are also rare in practice and could alter the replacement demand estimated by the stock-flow model. These mechanisms provide important directions for future work as more consistent GSE-specific data become available.

\subsection*{Interpreting the model and its limitations}

Several caveats are important when interpreting our results. First, the sequencing of shortages is shaped in part by the model structure. Because the lexicographic framework (detailed in Methods) prioritizes transformers over other supply-side and load-side GSE, the results should not be interpreted as showing that transformer adequacy is naturally more resilient in real-world markets. Rather, they show that the system cannot fully satisfy all GSE requirements simultaneously even when critical equipment is protected first. Real-world deployment may be more constrained still, because manufacturing lead times and plant-level throughput limits are not explicitly represented, so material availability may exceed what can actually be delivered as equipment within the planning horizon.

Second, the material constraints is a contributing factor to the GSE supply that this study uses as a proxy and does not capture a full set of conditions that determine equipment production. The BOM-derived material intensities characterize the physical material requirements of each GSE class, while the MRSUT represents the broader production and sourcing structure associated with those inputs, including intermediate production relationships, trade, and regional supply. Together, these layers capture both equipment-specific material needs and the upstream economic structure through which those materials reach electrical-equipment manufacturing. The observed sectoral flows therefore contain information about existing supply-chain capability, but do not directly measure equipment-specific manufacturing capacity. Plant-level assembly throughput, specialized component availability, skilled labor, qualification requirements, inventories, and equipment-specific production lead times may impose additional constraints on GSE supply \cite{nguyen2022electric, barlow2024analysis, baranowski2022wind}. The material representation is therefore most informative when the represented production and sourcing structure remains broadly applicable and upstream material flows are an important determinant of equipment availability.

Third, the MRSUT representation and the equipment-level BOM data are aggregated. They identify broad dependence patterns across sectors and regions, but do not fully resolve firm-level manufacturing pathways or design heterogeneity across voltage classes, applications, cooling configurations, manufacturers, and product generations, including technology innovation. Accordingly, the compiled BOM data should be interpreted as representative class-level material compositions, and the resulting BOM-derived material-intensity coefficients as representative modeling parameters rather than exact product specifications. The present framework focuses on primary material availability and equipment deployment. Secondary material supply from recycling is therefore not included in the material-availability constraints, while refurbishment and lifetime extension are not represented in the stock-flow formulation. Reuse, remanufacturing, and parts harvesting are also outside the current scope because they require component-level manufacturing and qualification data that are not consistently available across the GSE classes considered.

Fourth, the numerical results should be interpreted as scenario-based structural indicators rather than deterministic forecasts. The model does not yet endogenize prices, strategic behavior, inventories, substitution, learning, or adaptive procurement, all of which could alter the severity and structure of scarcity over time. Accordingly, the results are intended to characterize the potential scale, timing, and structure of GSE-related supply constraints under the modeled assumptions rather than to provide deterministic forecasts of future shortages.

\section*{METHODS}

\subsection*{Weibull-distributed survival-based modeling of GSE stock and replacement demand}
\label{Method:1}

To represent the evolution of installed GSE over time, we model each year's installations as a distinct cohort whose survival and replacement behavior follows an equipment-specific Weibull lifetime distribution \cite{li2002incorporating}. This cohort-based structure allows us to reconstruct historical stock accumulation from the beginning of the available record, quantify annual replacement demand, and project future in-service capacity with alternative lifetime assumptions.

For each GSE class, failure time is represented using a Weibull distribution with scale parameter $\alpha$ and shape parameter $\beta$. Let $F(t)$ denote the cumulative failure probability by age $t$, $S(t)$ the corresponding survival probability. Under an annual time step, the probability of failure during year $t$ is given by the increment in the cumulative failure probability:
\begin{align}
F(t) &= 1 - \exp\left[-\left(\frac{t}{\alpha}\right)^{\beta}\right], \\
S(t) &= \exp\left[-\left(\frac{t}{\alpha}\right)^{\beta}\right], \\
\Delta F(t) &= F(t) - F(t-1), \quad \Delta F(0)=0.
\end{align}
Here, $\Delta F(t)$ denotes the probability that a unit fails between ages $t-1$ and $t$. 

Let $C(y)$ denote the new GSE capacity required in year $y$ to support newly added generation or load-side capacity. This term represents new GSE demand driven by grid expansion and excludes replacement of previously installed equipment. Let $TC(y)$ denote the total GSE installation requirement in year $y$, including both this new capacity requirement and replacement demand arising from failures of previously installed cohorts.

For the first year of the record,
\begin{equation}
TC(1) = C(1).
\end{equation}
For each subsequent year $y \geq 2$, total installation is
\begin{equation}
TC(y) = C(y) + \sum_{k=1}^{y-1} TC(k)\,\Delta F(y-k),
\label{eq:TC}
\end{equation}
where the second term captures replacement demand from all prior cohorts. We use an annual discrete-time convention in which equipment installed in year $k$ enters the stock in that year. When observed in year $y>k$, the cohort has age $(y-k)$, and $\Delta F(y-k)=F(y-k)-F(y-k-1)$ gives the fraction of that cohort that retires during year $y$. The corresponding replacement requirement is assumed to be installed in the same year as the retirement.

The surviving capacity in target year $Y$ attributable to the cohort installed in year $y$ is
\begin{equation}
SC(y;Y) = TC(y)\,S(Y-y),
\label{eq:SC}
\end{equation}
and the total surviving stock in year $Y$ is
\begin{equation}
SC_{\text{total}}(Y) = \sum_{y=1}^{Y} TC(y)\,S(Y-y).
\end{equation}
Equations~\eqref{eq:TC} and \eqref{eq:SC} define an age-structured stock-and-flow formulation in which future installation needs are shaped jointly by net expansion and end-of-life replacement.

Weibull parameters were assigned separately for each GSE class using literature-based evidence on failure timing, wear-out behavior, and replacement dynamics, with full source details reported in the Supplemental Information. For transformers, the parameterization was informed by analytical and empirical studies on Weibull-based life estimation for power equipment, reliability modeling of large power transformers, and recent transformer-failure datasets developed for digitalized reliability analysis. For solar PV inverters, we relied on recent photovoltaic-system studies that treat inverter lifetime within the broader context of PV degradation, replacement, and long-run system performance. For wind power converters, including both DFIG and PMSG architectures, we used field-data studies of wind turbine fleets, including statistical failure analyses and component-level reliability assessments of converter systems. For data center UPS, we used reliability evidence for distributed UPS systems. For battery PCS, we drew on the stationary storage literature, including studies on battery system lifetime tradeoffs, reliability assessment of battery energy storage systems, and failure analysis of utility storage applications. For EV charger PCS, we used reliability analyses of fast-charging systems.

Because the available literature is heterogeneous in scope, technology definition, and reported failure statistics, we do not interpret any single source as definitive. Instead, we define two parameter sets for each equipment class: an \emph{optimistic} lifetime reflecting representative operating conditions, and a \emph{pessimistic} lifetime that shifts survival toward earlier retirement under more stressed operating environments, consistent with rising utilization and tighter operating margins under rapid load growth. These two parameterizations shown in Table~\ref{tab:weibull_parameters} provide bounded lifetime windows for the case study analysis and allow us to test the sensitivity of shortages to replacement timing. For data center transformers, we adopt the same Weibull parameters as for transformers because they are functionally transformer assets operating under similar core material and insulation constraints, while recognizing that their application environment may differ.

\begin{table}[t!]
\centering
\caption{\textbf{Weibull parameters used for stock survival and replacement modeling.}}
\label{tab:weibull_parameters}
\small
\setlength{\tabcolsep}{6pt}
\begin{tabular}{lcccc}
\toprule
\textbf{GSE class} & \multicolumn{2}{c}{\textbf{Pessimistic}} & \multicolumn{2}{c}{\textbf{Optimistic}} \\
\cmidrule(lr){2-3} \cmidrule(lr){4-5}
 & $\alpha$ & $\beta$ & $\alpha$ & $\beta$ \\
\midrule
Transformer & 40.9500 & 7.3410 & 49.5663 & 4.6141 \\
Data center transformer & 40.9500 & 7.3410 & 49.5663 & 4.6141 \\
SPV inverter & 16.2300 & 4.2300 & 25.5900 & 4.3500 \\
DFIG converter & 13.5000 & 0.6000 & 13.5000 & 0.7000 \\
PMSG converter & 30.3000 & 0.3400 & 30.3000 & 0.9500 \\
Data center UPS & 9.8100 & 2.8500 & 9.8100 & 2.8500 \\
EV charger PCS & 18.7800 & 5.0000 & 18.7800 & 5.0000 \\
Battery PCS & 19.5600 & 5.8300 & 38.4200 & 5.8600 \\
\bottomrule
\end{tabular}
\end{table}

The resulting parameterization captures substantial heterogeneity in survival behavior across GSE classes. Long-lived transformer assets exhibit later retirement profiles than power-electronic systems such as UPS, converter, inverter, and PCS equipment. This distinction is important because it determines whether future deployment is driven primarily by net expansion, replacement of aging legacy stock, or both.

\subsection*{Reconstruction of historical GSE stock and projection to 2030}
\label{Method:2}

To quantify the installed base and future deployment requirements of GSE, we reconstruct historical GSE capacity from observed generation and load development and extend these estimates to 2030 using scenario-based projections. The reconstruction begins from the earliest available historical record and converts primary generation or load capacity into associated GSE requirements using technology-specific capacity scaling ratios.

Historical generation capacity additions were derived from unit-level EIA-860 records \cite{eia860_data}, using the reported in-service year of each generator as the year of first deployment for each technology type. These generation additions were mapped to associated GSE requirements using engineering-based capacity multipliers that translate primary electrical capacity into required equipment capacity. The multipliers reflect the fact that GSE deployment generally exceeds nameplate generation or load capacity because of voltage transformation, redundancy, oversizing, and power conditioning requirements.

For power transformers, we apply capacity ratios of 1.1 for generator step-up (GSU) transformers, 2.25 for transmission transformers, and 2.32 for distribution transformers, representing the transformer capacity typically required at different stages of grid interconnection and delivery \cite{aubertin2016power}. For modeling purposes, we combine them into a single bulk transformer GSE category, giving a total multiplier of 5.67.

For inverter-based resources, we distinguish among major technology classes. Because EIA-860 does not distinguish between DFIG and PMSG wind technologies, the dataset does not directly specify whether turbines use partial-scale or full-scale converter configurations. We therefore infer converter topology from manufacturer, turbine model, commissioning period, and unit capacity. Models commonly associated with earlier geared fleets are classified as DFIG, whereas full-converter and direct-drive platforms are classified as PMSG. Based on this classification, historical wind additions are mapped to converter requirements using a converter-to-generator capacity ratio of 0.3 for DFIG systems and 1.0 for PMSG/full-converter systems. For forward projections, we assume that new wind deployment through 2030 is dominated by PMSG/full-converter configurations, whereas future DFIG-related GSE demand arises only from replacement of the existing installed base.

Utility-scale solar PV deployment is assigned an inverter loading ratio of 1.34 \cite{ramasamy2023us}, and battery energy storage systems (BESS) are assigned a PCS ratio of 1.0.

Because power grid datasets do not directly capture load-side GSE associated with data centers and EVs, these categories are estimated separately using external demand scenarios and engineering-based scaling ratios. For data centers, projected electrical demand is taken from the EPRI state-level dataset, with 2025 used as the reference year \cite{epri_powering_intelligence_2026}. Historical accumulation is reconstructed by assuming that large-scale commercial data center deployment begins in 2006 and increases linearly to the 2025 estimate. This reconstructed load is then mapped to GSE requirements using a transformer capacity ratio of 1.25 and a UPS-to-IT-load ratio of 1.37 \cite{paananen2023grid}. For EV charging, GSE requirements are estimated from external charging-demand scenarios \cite{wood20232030} and converted into charger-side PCS requirements using a 1:1 load-to-equipment mapping. Historical accumulation is reconstructed by treating 2011 as the initial year and assuming an exponential growth trajectory that scales installed charger-side PCS stock to the observed 2024 level.

All ratio definitions, classification rules, data sources, and calibration details are further provided in the Supplemental Information.

Future generation, data center, and EV-driven load growth are projected through 2030 using three exogenous demand components. General electricity demand and associated generation expansion follow the National Lab of the Rockies (NLR) standard scenarios under the ``Current Policies'' mid-case and high-demand-growth assumptions, corresponding to average electricity demand growth rates of 1.8\% and 2.8\% per year, respectively \cite{osti_2496240}. Data center growth follows medium- and high-growth trajectories defined by project development status: in the medium-growth case, all projects under construction, 75\% of projects in advanced planning, and 10\% of projects in early planning are assumed to be operational by 2030, whereas the high-growth case assumes full operation of all projects under construction and in advanced planning, together with 30\% of projects in early planning \cite{epri_powering_intelligence_2026}. EV growth follows mid- and high-adoption trajectories, corresponding to 33 million and 42 million EVs on the road by 2030, respectively \cite{wood20232030}. These scenario trajectories are mapped into GSE deployment requirements using the same technology-specific scaling ratios applied in the historical reconstruction, thereby preserving consistency between historical stock estimation and future equipment demand. Combined with the Weibull-based replacement model, this procedure yields annual GSE requirements through 2030 with both optimistic and pessimistic lifetime assumptions.

\subsection*{BOM compilation and derivation of GSE material intensity}
\label{Method:3}

To translate GSE capacity from electrical units into physical material demand, we compile representative equipment-level BOMs for the major supply- and load-side GSE classes considered in this study. For each equipment class, the BOM characterizes the mass and composition of constituent materials. We then normalize these material quantities by the rated apparent-power capacity of the equipment to derive material intensity coefficients in ton/GVA. These coefficients provide the quantitative link between GSE deployment and upstream material requirements and are used to convert reconstructed and projected GSE capacity into physical material demand.

The material accounting is defined over ten materials selected to capture both bulk materials that dominate GSE mass and lower-mass materials relevant to critical material and supply chain assessments reported by the U.S. Geological Survey (USGS) and the U.S. Department of Energy (DOE) \cite{usgs2025mcs, applegate2023final}. The resulting material set includes steel, copper, and aluminum as major bulk inputs, together with lower-mass but functionally important materials that appear in specific GSE architectures, including nickel, zinc, tin, silicon, silver, manganese, and magnesium.

Representative BOMs are developed for eight major GSE classes: conventional transformers, data center transformers, solar PV inverters, DFIG converters, PMSG converters, battery PCS, data center UPS, and EV charger PCS. These classes were selected to cover the dominant supply-side and load-side equipment categories analyzed in the case study.

Material composition data were compiled from multiple source types, including peer-reviewed literature, technical reports, teardown-style inventories, product documentation, and environmental product declarations (EPDs), with full source details reported in the Supplemental Information. For each source, material quantities were first harmonized to a common mass basis. Where the source reported an equipment- or component-level BOM, the material mass was then normalized by the corresponding rated apparent-power capacity. For material $m$ and equipment class $e$, the resulting material-intensity coefficient can be expressed as
\begin{equation}
    B_{m,e} = \frac{M_{m,e}}{K_e},
\end{equation}
where $M_{m,e}$ is the mass of material $m$ represented in the equipment-level BOM and $K_e$ is the corresponding rated equipment capacity. After unit harmonization, $B_{m,e}$ is expressed in ton/GVA. This common functional basis enables direct comparison of material requirements across equipment types with otherwise heterogeneous form factors and design bases.

When multiple sources were available for the same equipment class, we harmonized them to a common functional unit and derived representative average material intensity coefficients rather than adopting any single product specification. Thus, transformer BOMs were informed by transformer EPDs, inverter BOMs by photovoltaic inverter inventories and EPDs, battery PCS by storage-system and inverter documentation, UPS by data center UPS environmental profiles, and wind converter BOMs by renewable materials databases and converter-related technical documentation. Because GSE designs are not fully standardized and vary across manufacturers, voltage classes, cooling configurations, and product generations, these BOM values are interpreted as representative average compositions for each GSE class. Accordingly, the derived material intensity coefficients represent system-scale modeling parameters rather than the specification of any individual product. The aim is not to reproduce the exact material composition of any individual product, but to capture typical material requirements in a form suitable for system-scale modeling.

\subsection*{Tracing material supply chains using a multi-regional supply-use framework}
\label{Method:4}

To characterize the upstream material flows and regional sourcing associated with U.S. GSE manufacturing, we construct a multi-regional supply-use framework based on EXIOBASE. The BOM-derived material intensity coefficients developed in the preceding section characterize the equipment-side physical material requirements, while the MRSUT independently represents the production, trade, and regional sourcing structure associated with material supply to GSE-related manufacturing. These two components are linked later through the material feasibility constraint in the GSE deployment model.

We retain the native product-by-industry structure of the MRSUT, rather than starting from a pre-symmetrized input-output table, because this preserves the distinction between products and industries before imposing a technology assumption \cite{wood2014global}. This distinction is important in the present application, where GSE is not observed as a standalone EXIOBASE sector but must be isolated from a broader parent category.

We begin from the finest EXIOBASE resolution \cite{stadler_2025_15689391}, which contains 200 products and 163 industries. Within this system, Product 31 (\textit{Electrical machinery and apparatus n.e.c.}) and Industry 31 (\textit{Manufacture of electrical machinery and apparatus n.e.c.}) are treated as the parent product--industry category containing the major GSE classes considered here. This category provides the closest common EXIOBASE representation for transformers and electrical power-conversion equipment, while also containing electrical products outside the scope of this study. Because this parent category includes a broader set of electrical equipment beyond GSE itself, we downscale the associated sectoral flows using a global allocation factor of 4.6\%. This factor is derived from U.S. Annual Survey of Manufactures benchmark data by comparing the broader electrical-equipment manufacturing sector (NAICS 335) with the more GSE-relevant subcategories 335311 (Power, distribution, and specialty transformer manufacturing) and 335999 (All Other Miscellaneous Electrical Equipment and Component Manufacturing) \cite{census_asm_2018_2021}. The 4.6\% factor is used to approximate the GSE-relevant share of the broader electrical-machinery category represented in EXIOBASE.

To characterize uncertainty in this allocation, we apply the same sectoral mapping to Annual Survey of Manufactures data for each year from 2012 to 2021. The resulting GSE allocation share ranges from 4.15\% to 4.99\%, which we use as the historical lower and upper bounds in the sensitivity analysis. These bounds rescale the GSE-relevant EXIOBASE flows while leaving the BOM-derived material intensities, GSE demand, and regional sourcing structure otherwise unchanged.

We next identify upstream material categories associated with GSE production within the EXIOBASE framework \cite{stadler_2025_15689391}, focusing on metal and mineral product groups that capture the primary conductive, structural, and functional inputs to power-electronic and transformer equipment. These include aluminium, iron and steel, copper, precious metals, nickel, lead, zinc, tin, other non-ferrous metals, and non-metallic mineral products. Because each EXIOBASE category aggregates multiple underlying commodities, individual product groups may contain more than one material relevant to GSE manufacturing. We therefore further disaggregate these product groups using HS-code concordances to map EXIOBASE categories to the same material classes used in the material intensity analysis. Intermediate concordance tables are provided in the Supplementary Information.

Let $\mathbf{U}$ denote the multi-regional use table and $\mathbf{V}$ the corresponding supply table. Let $\hat{\mathbf{g}}$ and $\hat{\mathbf{q}}$ denote diagonal matrices of the total industry  and total product outputs. Following standard input-output notation \cite{miller2009input, wieland2018structural}, we define the normalized use matrix $\mathbf{B}$ and the market-share matrix $\mathbf{C}$ as
\begin{equation}
\mathbf{B} = \mathbf{U}\hat{\mathbf{g}}^{-1}, 
\quad
\mathbf{C} = \mathbf{V}\hat{\mathbf{q}}^{-1}.
\end{equation}

The matrix $\mathbf{B}$ describes the quantity of each product required per unit of industry output, whereas $\mathbf{C}$ describes the industry composition of supply for each product. Because these matrices are normalized by total output, their entries are coefficient values rather than monetary flow quantities. Neither matrix alone is sufficient to recover a product-by-product requirements system: $\mathbf{B}$ identifies how industries use products but not which industries produce a given product, while $\mathbf{C}$ identifies which industries supply a product but not the input structure associated with that production. A product-by-product representation therefore requires combining both relationships.

To construct this representation, we adopt the Industry Technology Assumption (ITA), under which each industry is assumed to produce all of its outputs using the same input structure \cite{miller2009input}. Under ITA, the input recipe is attached to the producing industry rather than to the individual product. The resulting symmetric product-by-product technical coefficients matrix is

\begin{equation}
\mathbf{A}^p = \mathbf{B}\mathbf{C}.
\end{equation}

The entries of $\mathbf{A}^p$ are dimensionless technical coefficients describing the upstream product inputs required per unit of product output. When applied to the GSE-relevant monetary activity represented within Product 31, this structure traces the corresponding direct and indirect upstream flows in monetary terms.

The resulting matrix $\mathbf{A}^p$ provides an average product-level requirements structure consistent with the observed product--industry and industry--product relationships in the MRSUT. Its main limitation is that it abstracts from within-industry heterogeneity across specific product lines. We therefore interpret the resulting sourcing structure as representative of GSE-related manufacturing patterns rather than specific to any individual firm or product.

To characterize the upstream production and sourcing structure associated with GSE manufacturing, we represent the layered supply chain of direct and indirect requirements generated by successive powers of $\mathbf{A}^p$. Under standard linear-production assumptions, where inputs are required in fixed proportions and production scales proportionally, this expansion is given by the Neumann series

\begin{equation}
(\mathbf{I} - \mathbf{A}^p)^{-1}
=
\sum_{k=0}^{\infty} (\mathbf{A}^p)^k,
\end{equation}
provided that the production system is productive, such that $\rho(\mathbf{A}^p)<1$ (i.e. total upstream requirements remain finite and the supply chain does not require more than one unit of output to produce itself) \cite{miller2009input}. In this representation, $(\mathbf{A}^p)^0 = \mathbf{I}$ denotes the unit of final demand, $(\mathbf{A}^p)^1$ captures direct upstream requirements, and higher-order terms $(\mathbf{A}^p)^k$ for $k \geq 2$ capture progressively more indirect supply chain layers.

In implementation, we evaluate this structure through a finite truncation
\begin{equation}
\sum_{k=0}^{n} (\mathbf{A}^p)^k,
\label{eq:truncated_neumann}
\end{equation}
which provides a layered approximation to the full embodied requirement system. This formulation is particularly useful in the present setting because Product 31 exhibits substantial intra-sectoral and cross-regional feedbacks. The matrix-power representation resolves these recursive interdependencies systematically and makes the layered structure of embodied sourcing explicit. The approximation error is given by the residual tail,
\begin{equation}
\mathbf{R}_n = \sum_{k=n+1}^{\infty} (\mathbf{A}^p)^k,
\end{equation}
which converges to zero as $n \rightarrow \infty$ when $\rho(\mathbf{A}^p)<1$. Accordingly, the truncated expansion in Equation~\eqref{eq:truncated_neumann} provides a consistent approximation framework.

This layered decomposition is interpreted under the standard assumptions of input-output analysis, including constant returns to scale and time-invariant technical coefficients over the modeled period \cite{ten2005economics}. Formally, each coefficient $a_{ij}^{rs}\in\mathbf{A}^p$, representing the intermediate requirement of product $i$ from region $r$ per unit of output of product $j$ in region $s$, is assumed to be independent of output volume. Within the EXIOBASE framework, this further implies destination independence of product structure: the embodied input recipe is determined by the producing region and sector rather than by the destination to which the product is delivered \cite{stadler2018exiobase}. Under these assumptions, the same technical structure applies across successive upstream layers, allowing GSE manufacturing upstream material flows to be traced consistently through the multi-regional system.

The MRSUT analysis provides GSE-related upstream material flows in monetary terms. To express these flows in physical units, we map the selected EXIOBASE material categories to corresponding Harmonized System (HS) codes and use UN Comtrade \cite{UNComtrade2026} data on trade value and mass to calculate the tons of each material per unit of monetary value. For each material, the conversion can be written dimensionally as
\[
[\text{monetary flow}]
\times
\left[\frac{\text{ton}}{\text{monetary value}}\right]
=
[\text{ton}].
\]
Using this conversion, the regional MRSUT flows can be expressed in tons. We denote the resulting material availability associated with GSE-related manufacturing from region $r$ for material $m$ in year $y$ as $A_{r,m,y}$.

This conversion puts the MRSUT results and the equipment material requirements in the same physical unit. The BOM-derived material intensity $B_{m,e}$ is expressed in ton/GVA, and deployed GSE capacity $P_{e,y}$ is expressed in GVA, so that
\[
\left[\frac{\text{ton}}{\text{GVA}}\right]
\times
[\text{GVA}]
=
[\text{ton}].
\]
Therefore, $B_{m,e}P_{e,y}$ gives the amount of material required for GSE deployment in tons, while $A_{r,m,y}$ gives the available amount of that material in tons. Equation~\eqref{eq:material_limit} in the deployment model presented in the next subsection then directly compares the required and available amounts.

The resulting regional material flows are used in the Results section to assess regional concentration, import dependence, and exposure to trade disruption. Intermediate and processed data are provided in the Supplementary Data.

\subsection*{Top-down GSE deployment model}
\label{sec:methodology_top_down}

To identify deployment gaps under material supply constraints, we formulate a top-down optimization model that allocates available materials across GSE categories in the target study region. The framework is designed to reflect two system features. First, deployment depends on aggregate upstream material availability rather than on unconstrained equipment supply. Second, different GSE classes do not contribute symmetrically to system functionality, so their deployment is prioritized hierarchically. In particular, critical grid-enabling equipment is protected before generation-side and load-side equipment. The model therefore identifies how much of the projected GSE demand can be met in each year, and where unmet demand emerges once material constraints become binding. Accordingly, the timing and distribution of shortages across GSE classes reflect both their underlying demand and material requirements and the allocation structure imposed under scarcity.

The optimization is based on a lexicographic expansion logic. Rather than allowing the model to allocate scarce materials to only one high-priority category, we first maximize the deployment of proportionally complete equipment bundles across grid, generation, and consumption layers. This represents the idea that useful system expansion often requires coordinated deployment across multiple equipment categories. When fully proportioned bundles can no longer be completed, the model then falls back to a strict hierarchy that prioritizes critical grid-enabling equipment.

The sets and indices are defined as follows:
\begin{itemize}
    \item $\mathcal{R}$: set of material supply regions, indexed by $r$.
    \item $\mathcal{M}$: set of critical materials, indexed by $m$.
    \item $\mathcal{Y}$: set of years, indexed by $y$.
    \item $\mathcal{E}$: set of all GSE classes, indexed by $e$.
    \item $\mathcal{E}_{\text{grid}} \subset \mathcal{E}$: critical grid-enabling equipment (i.e., transformer).
    \item $\mathcal{E}_{\text{gen}} \subset \mathcal{E}$: generation-side equipment (i.e., SPV inverter, DFIG converter, PMSG converter, battery PCS).
    \item $\mathcal{E}_{\text{cons}} \subset \mathcal{E}$: load-side equipment (i.e., data center transformer, data center UPS, EV charger PCS).
\end{itemize}

The model uses the following parameters:
\begin{itemize}
    \item $A_{r,m,y}$: available mass of material $m$ from supply region $r$ in year $y$ (ton).
    
    \item $D_{e,y}$: target demand for equipment class $e$ in year $y$ (GVA).
    
    \item $B_{m,e}$: BOM-derived material intensity coefficient for material $m$ and GSE class $e$, expressed as the mass of material required per unit of equipment capacity (ton/GVA).
    
    \item $\rho_e$: capacity requirement ratio for GSE class $e$, defined as the required GSE capacity per unit of associated generation or load growth (e.g., 5.67 for transformers, 1.34 for SPV inverters, 0.3 for DFIG converters, 1.0 for PMSG converters, 1.0 for battery PCS, 1.25 for data center transformers, 1.37 for data center UPS, and 1.0 for EV charger PCS).
    
    \item $w_e$: priority weight assigned to equipment class $e$, with
    \[
    w_{e \in \mathcal{E}_{\text{grid}}} \gg
    w_{e \in \mathcal{E}_{\text{gen}}} \gg
    w_{e \in \mathcal{E}_{\text{cons}}}.
    \]
    
    \item $M$: a sufficiently large scalar satisfying $M \gg w_e$ for all $e \in \mathcal{E}$, used to enforce lexicographic priority of synchronous bundle formation.
\end{itemize}

The decision variables are:
\begin{itemize}
    \item $P_{e,y} \geq 0$: deployed production capacity of equipment class $e$ in year $y$.
    \item $U_{e,y} \geq 0$: unmet demand for equipment class $e$ in year $y$.
    \item $V_{e,y} \geq 0$: normalized deployment level of equipment class $e$ in year $y$.
    \item $S_y \geq 0$: synchronous bundle level in year $y$, representing the maximum number of proportionally complete equipment sets that can be deployed simultaneously.
\end{itemize}

The objective function is
\begin{equation}
\label{eq:objective_lex}
\max \sum_{y \in \mathcal{Y}}
\left(
M \cdot S_y + \sum_{e \in \mathcal{E}} w_e V_{e,y}
\right).
\end{equation}
The first term maximizes synchronous bundle deployment and therefore enforces coordinated system expansion across equipment categories. The second term allocates any remaining feasible deployment according to the hierarchical priority structure.

Material feasibility is enforced by requiring that total material consumption across all deployed GSE classes not exceed the aggregate material supply available from all sourcing regions:
\begin{equation}
\label{eq:material_limit}
\sum_{e \in \mathcal{E}} B_{m,e} P_{e,y}
\leq
\sum_{r \in \mathcal{R}} A_{r,m,y}
\qquad
\forall m \in \mathcal{M}, \forall y \in \mathcal{Y}.
\end{equation}
Here, $A_{r,m,y}$ represents the primary material supply available from region $r$ and does not include an endogenous contribution from recycling or end-of-life GSE recovery. Component-level circularity pathways, including reuse, remanufacturing, and parts harvesting, are also outside the model scope because GSE supply is represented through material availability rather than component-level manufacturing flows.

For each equipment class and year, deployed capacity and unmet demand must sum to the projected requirement:
\begin{equation}
\label{eq:demand_balance}
P_{e,y} + U_{e,y} = D_{e,y}
\qquad
\forall e \in \mathcal{E}, \forall y \in \mathcal{Y}.
\end{equation}

To compare heterogeneous equipment classes within a common synchronous-bundle framework, absolute deployment is converted into normalized deployment units using the capacity requirement ratio:
\begin{equation}
\label{eq:norm_prod}
V_{e,y} = \frac{P_{e,y}}{\rho_e}
\qquad
\forall e \in \mathcal{E}, \forall y \in \mathcal{Y}.
\end{equation}
The synchronous bundle level is then constrained by the least available normalized deployment among all active equipment classes:
\begin{equation}
\label{eq:sync_bundle}
S_y \leq V_{e,y}
\qquad
\forall e \in \mathcal{E} \text{ with } D_{e,y} > 0,\ \forall y \in \mathcal{Y}.
\end{equation}
This formulation ensures that $S_y$ represents the number of fully proportioned bundles that can be deployed simultaneously in year $y$.

To preserve the top-down hierarchy when synchronous bundle deployment cannot be fully maintained, we impose an upper bound linking all non-grid equipment to the normalized deployment of critical grid equipment. Let $e^* \in \mathcal{E}_{\text{grid}}$ denote the reference grid-enabling equipment class. Then
\begin{equation}
\label{eq:hierarchy_bound}
V_{e,y} \leq V_{e^*,y}
\qquad
\forall e \in \mathcal{E} \setminus \{e^*\}, \forall y \in \mathcal{Y}.
\end{equation}
This prevents the model from expanding generation-side or load-side equipment beyond the scale supported by critical grid infrastructure. This hierarchy is an allocation assumption under material scarcity rather than an empirical manufacturing priority. Once proportional deployment across all GSE classes can no longer be maintained, it tends to preserve transformer deployment relative to generation-side and load-side GSE. The later appearance of transformer shortages in the Results therefore reflects, in part, this allocation structure in addition to differences in equipment demand, replacement requirements, and material intensity.

Taken together, Equations~\eqref{eq:objective_lex}--\eqref{eq:hierarchy_bound} define a constrained deployment problem in which projected equipment demand, replacement requirements, material intensity and availability, equipment proportionality, and the allocation hierarchy jointly determine realized GSE deployment and unmet demand. The resulting solution provides the basis for identifying when and where material supply limitations translate into equipment shortages across the study horizon.

\newpage

\section*{RESOURCE AVAILABILITY}

\subsection*{Lead contact}

Further information and requests for resources should be directed to and will be fulfilled by the lead contact, Yury Dvorkin (ydvorki1@jhu.edu).

\subsection*{Materials availability}

This study did not generate new materials.

\subsection*{Data and code availability}

\begin{itemize}
    \item The data generated and processed in this study, along with the code used for the analysis, are described in the Supplementary Information and are available on Zenodo: \url{https://doi.org/10.5281/zenodo.19597635}.
    \item Additional information required to reanalyze the data reported in this study is available from the lead contact upon reasonable request.
\end{itemize}

\section*{ACKNOWLEDGMENTS}

This work was supported in part by the Johns Hopkins University Discovery Award and in part by the U.S. National Science Foundation under Grant OISE-2330450.

\section*{AUTHOR CONTRIBUTIONS}

Conceptualization, B.Y. and Y.D.; methodology, B.Y. and Y.D.; investigation, B.Y. and Y.D.; visualization, B.Y.; writing–original draft, B.Y.; writing–review \& editing, B.Y. and Y.D.; funding acquisition, Y.D.; resources, B.Y. and Y.D.; supervision, Y.D.

\section*{DECLARATION OF INTERESTS}

The authors declare no competing interests.

\section*{DECLARATION OF GENERATIVE AI AND AI-ASSISTED TECHNOLOGIES}

During the preparation of this work, the authors used ChatGPT to assist with grammar checking and \LaTeX{} formatting. After using this tool, the authors reviewed and edited the content as needed and take full responsibility for the content of the publication.

\newpage

% \section*{MAIN FIGURE TITLES AND LEGENDS}

% \newpage

% \section*{MAIN TABLES, INCLUDING TITLES AND LEGENDS}

\bibliography{ref_intro, ref_case_study, ref_current_electronics, ref_EPD, ref_methodology, ref_methodology_material, ref_methodology_sut, ref_optimization, ref_weibull, ref_discussion}

\end{document}